\documentclass[twocolumn,showpacs,preprintnumbers,amsmath,amssymb]{revtex4}

\usepackage{graphicx}
\usepackage{dcolumn}
\usepackage{bm}

\begin{document}

\title{Structure and magnetism of self-organized Ge$_{1-x}$Mn$_{x}$ nano-columns}

\author{~T. Devillers}
 \email{thibaut.devillers@cea.fr}
 \affiliation{CEA-Grenoble/DSM/DRFMC/SP2M,\\ 17 rue des Martyrs,\\ 38054 Grenoble Cedex 9, France}%
\author{M. Jamet}%
 \email{matthieu.jamet@cea.fr}
 \affiliation{CEA-Grenoble/DSM/DRFMC/SP2M,\\ 17 rue des Martyrs,\\ 38054 Grenoble Cedex 9, France}%
\author{A. Barski}
\affiliation{CEA-Grenoble/DSM/DRFMC/SP2M,\\ 17 rue des Martyrs,\\ 38054 Grenoble Cedex 9, France}%
\author{V. Poydenot}
\affiliation{CEA-Grenoble/DSM/DRFMC/SP2M,\\ 17 rue des Martyrs,\\ 38054 Grenoble Cedex 9, France}%
\author{P. Bayle-Guillemaud}
\affiliation{CEA-Grenoble/DSM/DRFMC/SP2M,\\ 17 rue des Martyrs,\\ 38054 Grenoble Cedex 9, France}%
\author{E. Bellet-Amalric}
\affiliation{CEA-Grenoble/DSM/DRFMC/SP2M,\\ 17 rue des Martyrs,\\ 38054 Grenoble Cedex 9, France}%

\author{S. Cherifi}
\affiliation{Institut N\'eel,\\ CNRS, BP166,\\ 38042 Grenoble Cedex 9, France.}
\author{J. Cibert}
\affiliation{Institut N\'eel,\\ CNRS, BP166,\\ 38042 Grenoble Cedex 9, France.}

\date{\today}

\begin{abstract}
We report on the structural and magnetic properties of thin Ge$_{1-x}$Mn$_{x}$ films grown by molecular beam epitaxy (MBE) on Ge(001) substrates at temperatures ($T_{g}$) ranging from $80^{\circ}$C to 200$^{\circ}$C, with average Mn contents between 1 \% and 11 \%. Their crystalline structure, morphology and composition have been investigated by transmission electron microscopy (TEM), electron energy loss spectroscopy and x-ray diffraction. In the whole range of growth temperatures and Mn concentrations, we observed the formation of manganese rich nanostructures embedded in a nearly pure germanium matrix. Growth temperature mostly determines the structural properties of Mn-rich nanostructures. For low growth temperatures (below 120$^{\circ}$C), we evidenced a two-dimensional spinodal decomposition resulting in the formation of vertical one-dimensional nanostructures (nanocolumns). Moreover we show in this paper the influence of growth parameters ($T_{g}$ and Mn content) on this decomposition \textit{i.e.} on nanocolumns size and density. For temperatures higher than $180^{\circ}$C, we observed the formation of Ge$_3$Mn$_5$ clusters. For intermediate growth temperatures nanocolumns and nanoclusters coexist.
Combining high resolution TEM and superconducting quantum interference device magnetometry, we could evidence at least four different magnetic phases in Ge$_{1-x}$Mn$_{x}$ films: $(i)$ paramagnetic diluted Mn atoms in the germanium matrix, $(ii)$ superparamagnetic and ferromagnetic low-$T_{C}$ nanocolumns (120 K $\leq$ $T_{C}$ $\leq$ 170 K), $(iii)$ high-$T_{C}$ nanocolumns ($T_{C}$ $\geq$ 400 K) and $(iv)$ Ge$_{3}$Mn$_{5}$ clusters. 

\end{abstract}

\pacs{75.50.Pp, 75.75.+a, 61.46.-w}
\maketitle

\section{introduction}
In the past few years, the synthesis of ferromagnetic semiconductors has become a major challenge for spintronics. Actually, growing a magnetic and semiconducting material could lead to promising advances like spin injection into non magnetic semiconductors, or electrical manipulation of carrier induced magnetism in magnetic semiconductors \cite{ohno00,Bouk02}. Up to now, major efforts have focused on diluted magnetic semiconductors (DMS) in which the host semiconducting matrix is randomly substituted by transition metal (TM) ions such as Mn, Cr, Ni, Fe or Co \cite{Diet02}. However Curie temperatures ($T_{C}$) in DMS remain rather low and TM concentrations must be drastically raised in order to increase $T_{C}$ up to room temperature. That usually leads to phase separation and the formation of secondary phases. It was recently shown that phase separation induced by spinodal decomposition could lead to a significant increase of $T_{C}$ \cite{Diet06,Fuku06}. For semiconductors showing $T_{C}$ higher than room temperature one can foresee the fabrication of nanodevices such as memory nanodots, or nanochannels for spin injection. Therefore, the precise control of inhomogeneities appears as a new challenge which may open a way to industrial applications of ferromagnetism in semiconductors.

The increasing interest in group-IV magnetic semiconductors can also be explained by their potential compatibility with the existing silicon technology. In 2002, carrier mediated ferromagnetism was reported in  MBE grown Ge$_{0.94}$Mn$_{0.06}$ films by Park \textit{et al.} \cite{Park02}. The maximum critical temperature was 116 K. Recently many publications indicate a significant increase of $T_{C}$ in Ge$_{1-x}$Mn$_{x}$ material depending on growth conditions \cite{Pint05,Li05,tsui03}. Cho \textit{et al.} reported a Curie temperature as high as 285 K \cite{Cho02}. 
Taking into account the strong tendency of Mn ions to form intermetallic compounds in germanium, a detailed investigation of the nanoscale structure is required. Up to now, only a few studies have focused on the nanoscale composition in Ge$_{1-x}$Mn$_{x}$ films. Local chemical inhomogeneities have been recently reported by Kang \textit{et al.} \cite{Kang05} who evidenced a micrometer scale segregation of manganese in large Mn rich stripes. Ge$_3$Mn$_5$ as well as Ge$_8$Mn$_{11}$ clusters embedded in a germanium matrix have been reported by many authors. However, Curie temperatures never exceed 300 K \cite{Bihl06,Morr06,Pass06,Ahle06}. Ge$_3$Mn$_5$ clusters exhibit a Curie temperature of 296 K \cite{Mass90}. This phase frequently observed in Ge$_{1-x}$Mn$_{x}$ films is the most stable (Ge,Mn) alloy. The other stable compound Ge$_8$Mn$_{11}$ has also been observed in nanocrystallites surrounded with pure germanium \cite{Park01}. Ge$_8$Mn$_{11}$ and Ge$_3$Mn$_5$ phases are ferromagnetic but their metallic character considerably complicates their potential use as spin injectors.
Recently, some new Mn-rich nanostructures have been evidenced in Ge$_{1-x}$Mn$_{x}$ layers. Sugahara \textit{et al.} \cite{Sugh05} reported the formation of high Mn content (between 10 \% and 20 \% of Mn) amorphous Ge$_{1-x}$Mn$_x$ precipitates  in a Mn-free germanium matrix. Mn-rich coherent cubic clusters were observed by Ahlers \textit{et al.} \cite{Ahle06} which exhibit a Curie temperatures below 200 K. Finally, high-$T_{C}$ ($>$ 400 K) Mn-rich nanocolumns have been evidenced \cite{Jame06} which could lead to silicon compatible room temperature operational devices.\\
In the present paper, we investigate the structural and magnetic properties of Ge$_{1-x}$Mn$_x$ thin films for low growth temperatures ($<$ 200$^{\circ}$C) and low Mn concentrations (between 1 \% and 11 \%). By combining TEM, x-Ray diffraction and SQUID magnetometry, we could identify different magnetic phases. We show that depending on growth conditions, we obtain either Mn-rich nanocolumns or Ge$_{3}$Mn$_{5}$ clusters embedded in a germanium matrix. We discuss the structural and magnetic properties of these nanostructures as a function of manganese concentration and growth temperature. We also discuss the magnetic anisotropy of nanocolumns and  
Ge$_3$Mn$_5$ clusters. 

\section{Sample growth}

Growth was performed using solid sources molecular beam epitaxy (MBE) by co-depositing Ge and Mn evaporated from standard Knudsen effusion cells.  Deposition rate was low ($\approx$ 0.2 \AA.s$^{-1}$). Germanium substrates were epi-ready Ge(001) wafers with a residual n-type doping and resistivity of 10$^{15}$ cm$^{-3}$ and 5 $\Omega.cm$ respectively. After thermal desorption of the surface oxide, a 40 nm thick Ge buffer layer was grown at 250$^{\circ}$C, resulting in a 2 $\times$ 1 surface reconstruction as observed by reflection high energy electron diffraction (RHEED) (see Fig. 1a). Next, 80 nm thick Ge$_{1-x}$Mn$_{x}$ films were subsequently grown  at low substrate temperature (from 80$^{\circ}$C to 200$^{\circ}$C). Mn content has been determined by x-ray fluorescence measurements performed on thick samples ($\approx$ 1 $\mu m$ thick) and complementary Rutherford Back Scattering (RBS) on thin Ge$_{1-x}$Mn$_{x}$ films grown on silicon. Mn concentrations range from 1 \% to 11\% Mn.

For Ge$_{1-x}$Mn$_{x}$ films grown at substrate temperatures below 180$^{\circ}$C, after the first monolayer (ML) deposition, the  2 $\times$ 1 surface reconstruction almost totally disappears. After depositing few MLs, a slightly diffuse 1 $\times$ 1 streaky RHEED pattern and a very weak 2 $\times$ 1 reconstruction (Fig. 1b) indicate a predominantly two-dimensional growth. For growth temperatures above 180$^{\circ}$C additional spots appear in the RHEED pattern during the Ge$_{1-x}$Mn$_{x}$ growth (Fig. 1c). These spots may correspond to the formation of very small secondary phase crystallites. The nature of these crystallites will be discussed below.

Transmission electron microscopy (TEM) observations were performed using a JEOL 4000EX microscope with an acceleration voltage of 400 kV. Energy filtered transmission electron microscopy (EFTEM) was done using a JEOL 3010 microscope equipped with a Gatan Image Filter . Sample preparation was carried out by standard mechanical polishing and argon ion milling for cross-section investigations and plane views were prepared by wet etching with H$_3$PO$_4$-H$_2$O$_2$ solution \cite{Kaga82}.

\begin{figure}[htb]
    \center
    \includegraphics[width=.29\linewidth]{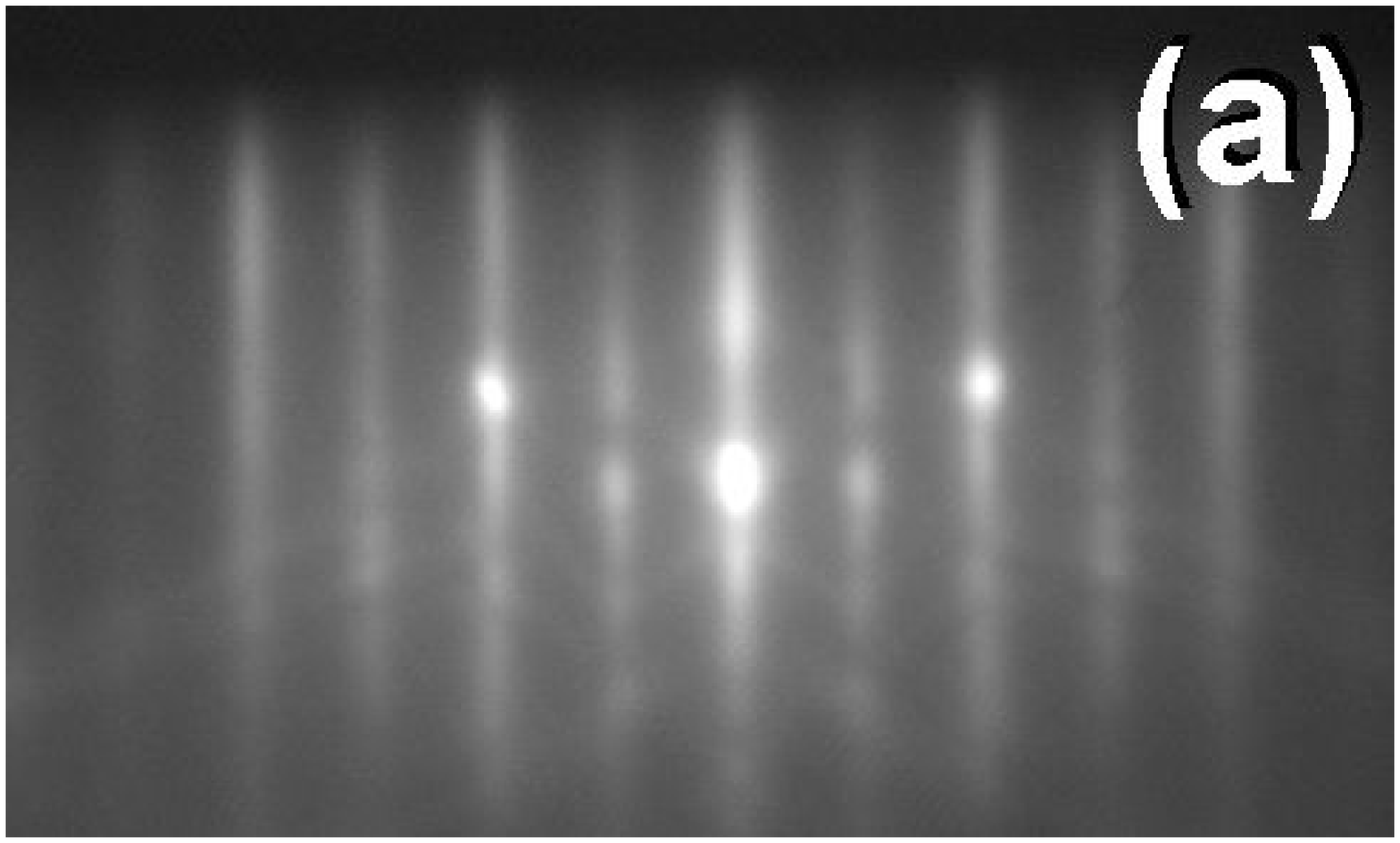}
    \includegraphics[width=.29\linewidth]{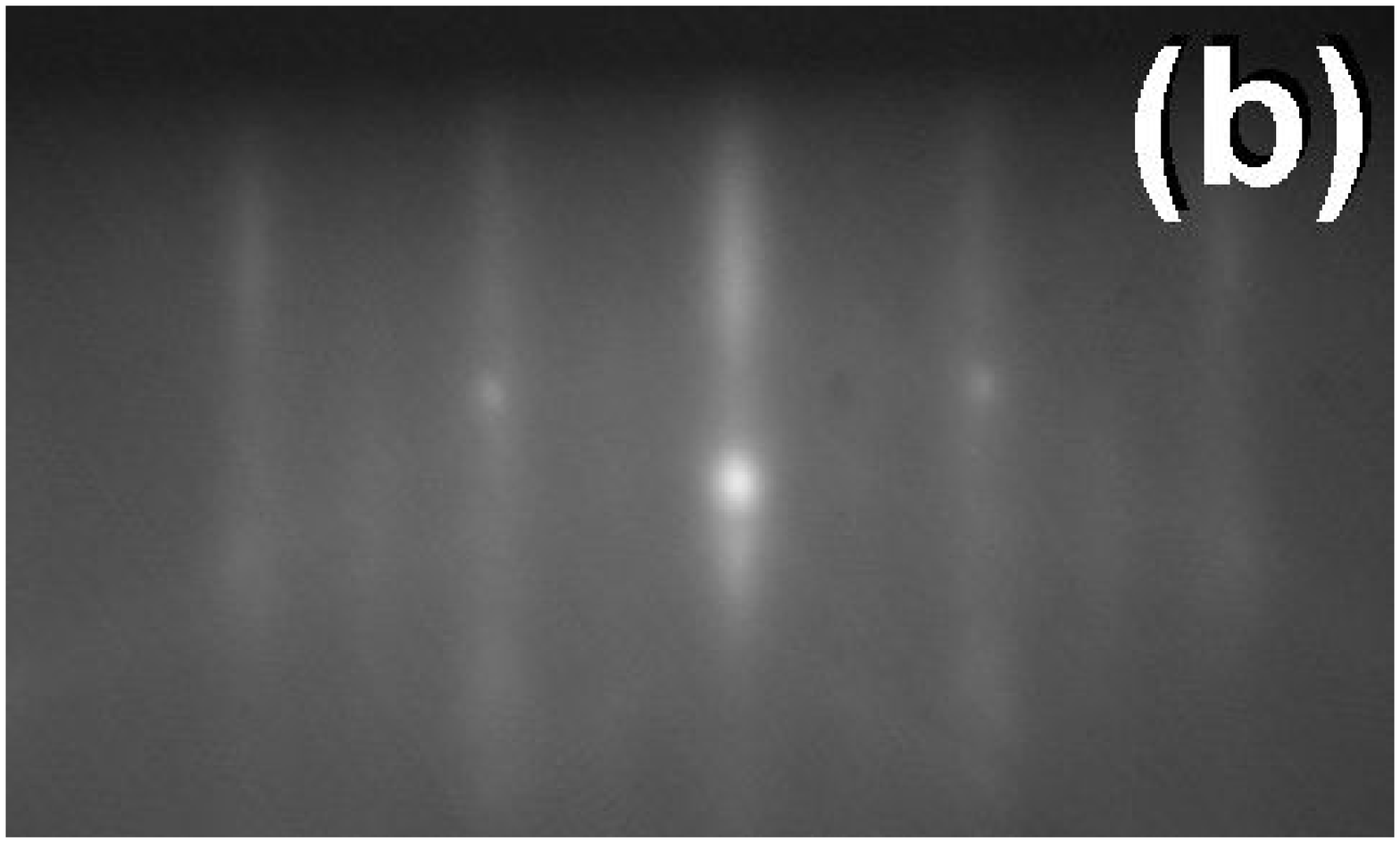}
    \includegraphics[width=.29\linewidth]{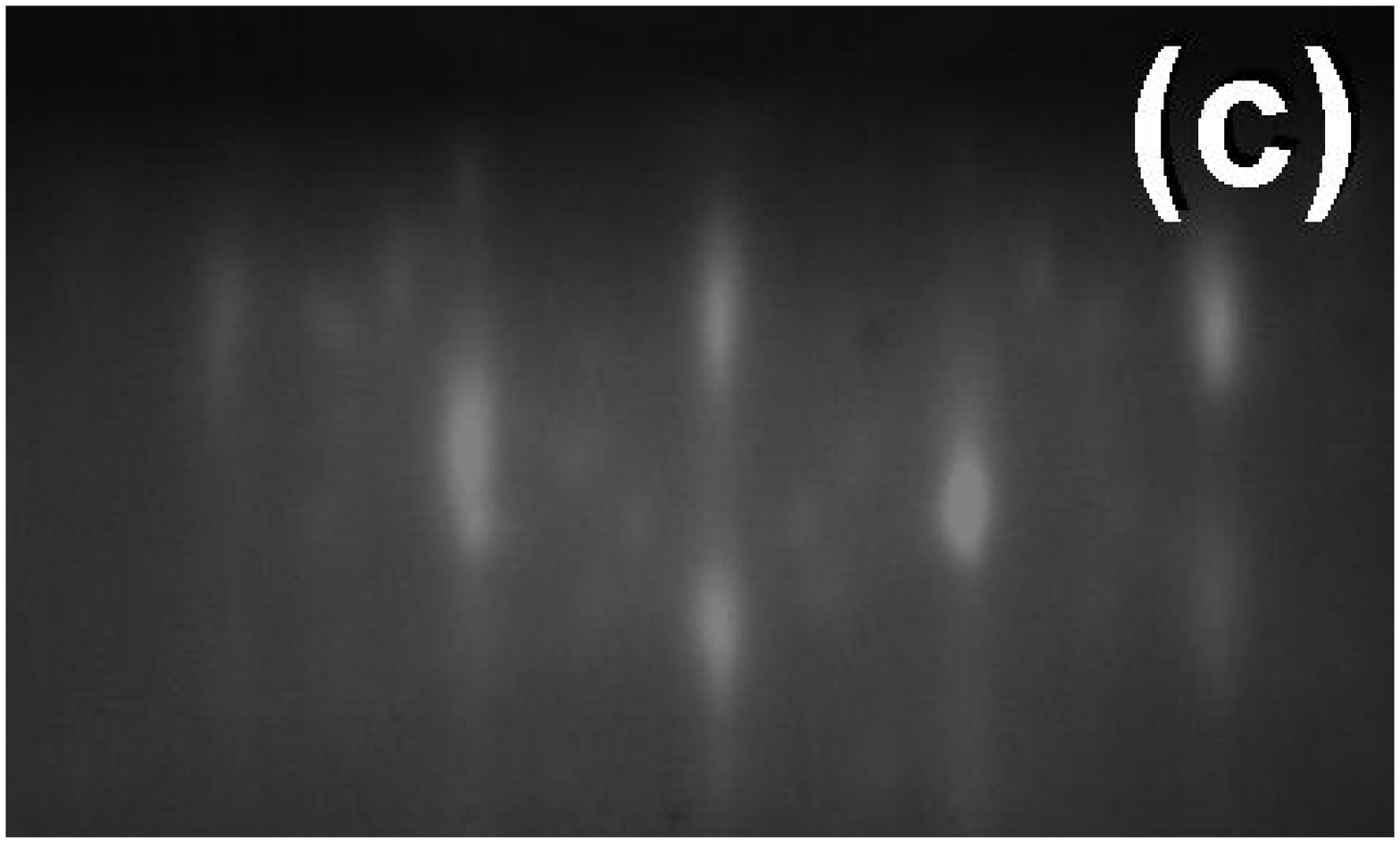}
    \caption{RHEED patterns recorded during the growth of Ge$_{1-x}$Mn$_{x}$ films: (a) 2 $\times$ 1 surface reconstruction of the germanium buffer layer. (b) 1 $\times$ 1 streaky RHEED pattern obtained at low growth temperatures ($T_g<$180$^{\circ}$C). (c) RHEED pattern of a sample grown at $T_g=$180$^{\circ}$C. The additional spots reveal the presence of Ge$_3$Mn$_5$ clusters at the surface of the film.}
\label{fig1}
\end{figure}

\section{Structural properties \label{structural}}

\begin{figure}[htb]
    \center
	\includegraphics[width=.49\linewidth]{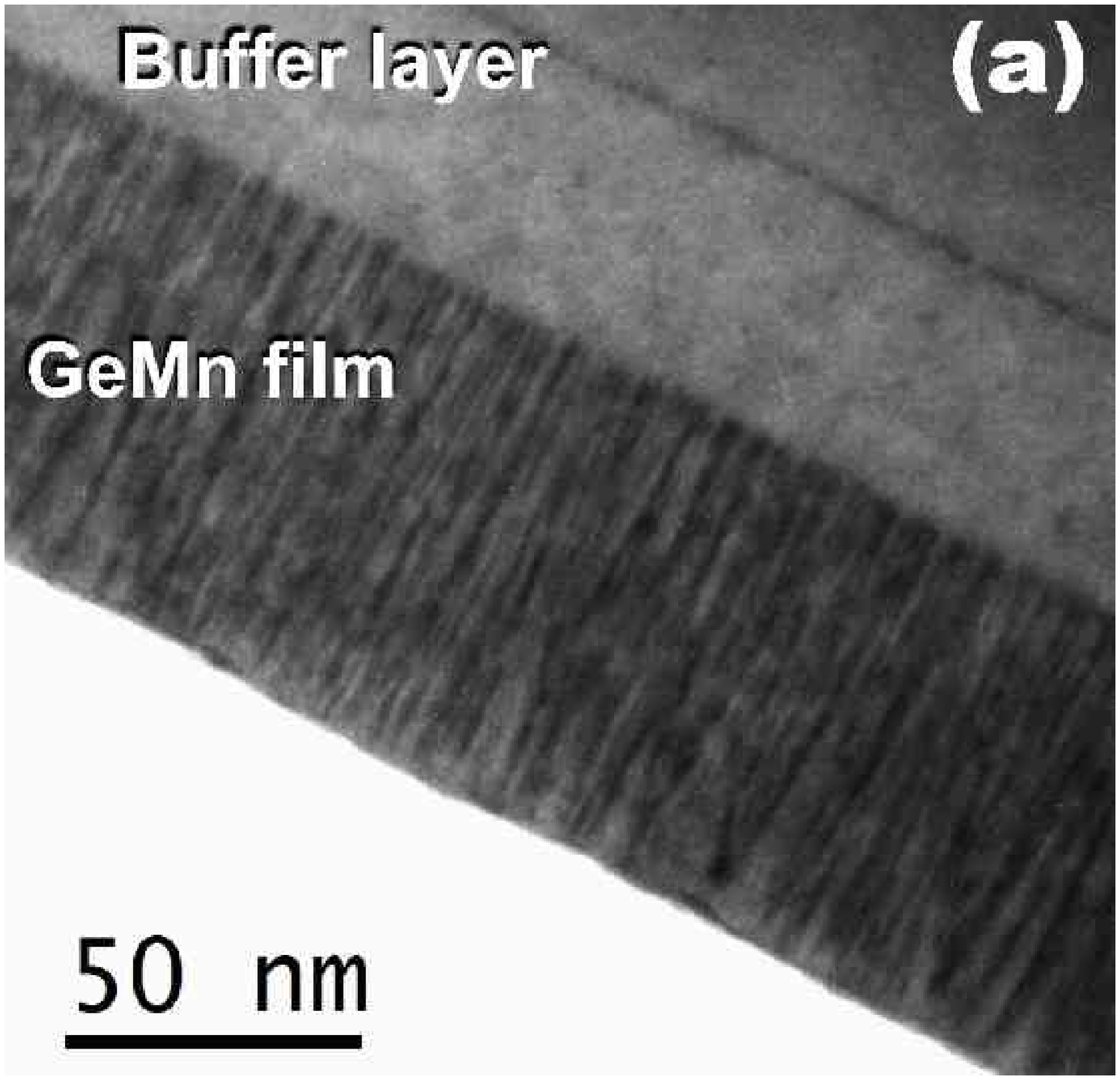}
	\includegraphics[width=.49\linewidth]{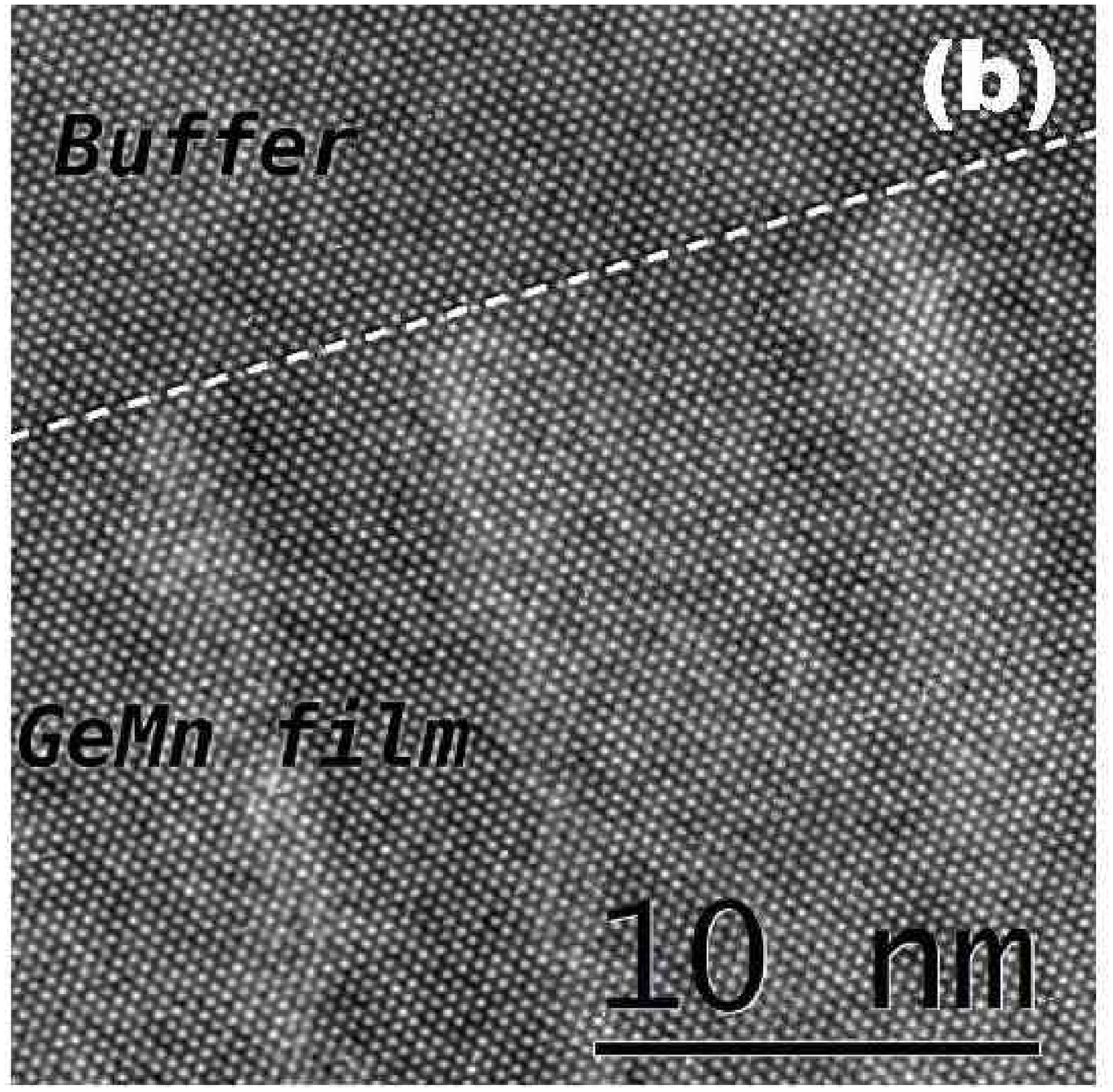}
	 \includegraphics[width=.49\linewidth]{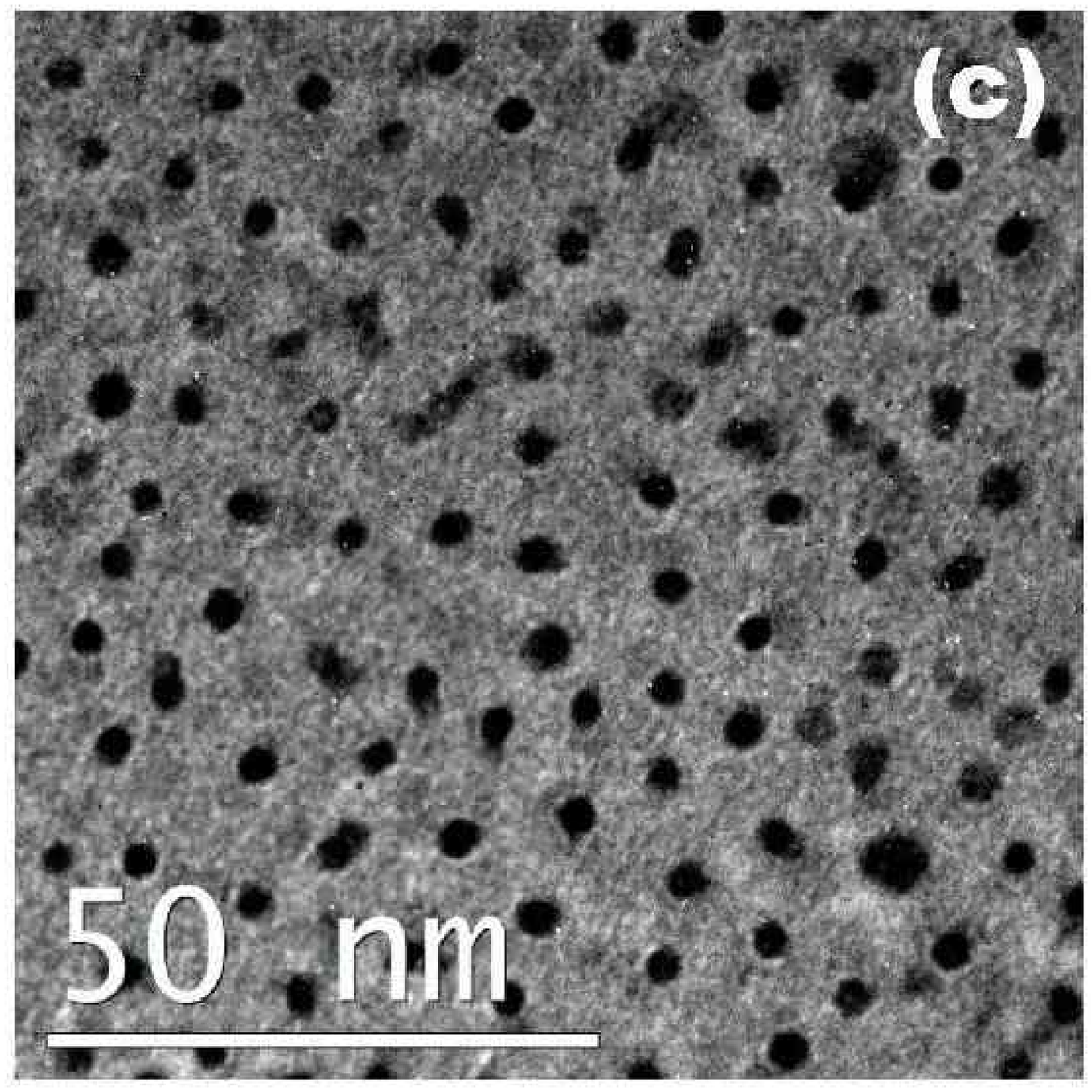}
	 \includegraphics[width=.49\linewidth]{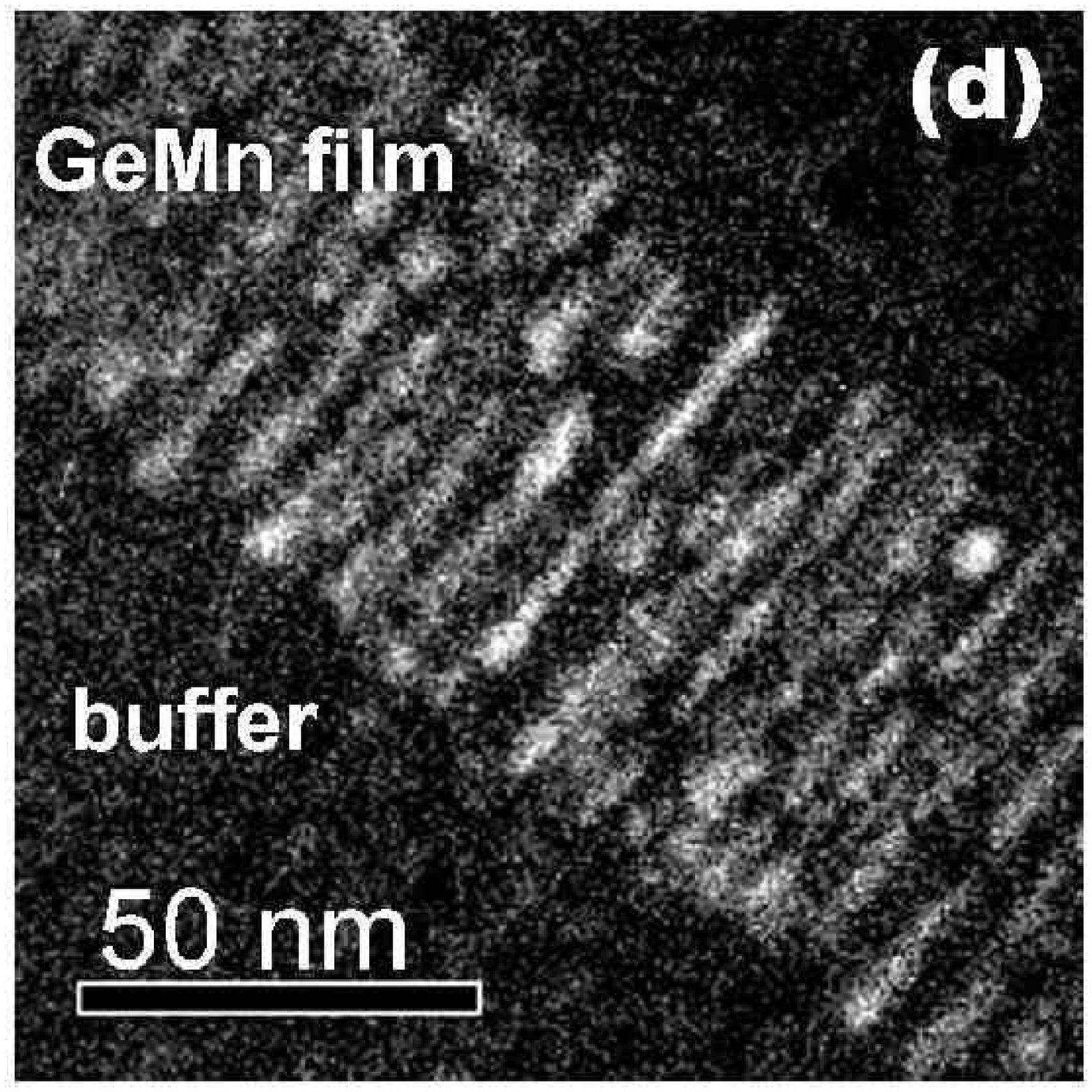}
    \caption{Transmission electron micrographs of a Ge$_{1-x}$Mn$_{x}$ film grown at 130$^{\circ}$C and containing 6 \% of manganese. (a) cross-section along the [110] axis : we clearly see the presence of nanocolumns elongated along the growth axis. (b) High resolution image of the interface between the Ge$_{1-x}$Mn$_{x}$ film and the Ge buffer layer. The Ge$_{1-x}$Mn$_{x}$ film exhibits the same diamond structure as pure germanium. No defect can be seen which could be caused by the presence of nanocolumns. (c) Plane view micrograph performed on the same sample confirms the columnar structure and gives the density and size distribution of nanocolumns. (d) Mn chemical map obtained by energy filtered transmission electron microcopy (EFTEM). The background was carefully substracted from pre-edge images. Bright areas correspond to Mn-rich regions.}
\label{fig2}
\end{figure}

In samples grown at 130$^{\circ}$C and containing 6 \% Mn, we can observe vertical elongated nanostructures \textit{i.e.} nanocolumns as shown in Fig. 2a. Nanocolumns extend through the whole Ge$_{1-x}$Mn$_{x}$ film thickness. From the high resolution TEM image shown in Fig. 2b, we deduce their average diameter around 3 nm. Moreover in Fig. 2b, the interface between the Ge buffer layer and the Ge$_{1-x}$Mn$_{x}$ film is flat and no defect propagates from the interface into the film. The Ge$_{1-x}$Mn$_{x}$ film is a perfect single crystal in epitaxial relationship with the substrate. In Fig. 2c is shown a plane view micrograph of the same sample confirming the presence of nanocolumns in the film. From this image, we can deduce the size and density of nanocolumns. The nanocolumns density is 13000 $\rm{\mu m}^{-2}$ with a mean diameter of 3 nm which is coherent with cross-section measurements. In order to estimate the chemical composition of these nanocolumns, we further performed chemical mapping using EFTEM. In Fig. 2d we show a cross sectional Mn chemical map of the Ge$_{1-x}$Mn$_{x}$ film. This map shows that the formation of nanocolumns is a consequence of Mn segregation. Nanocolumns are Mn rich and the surrounding matrix is Mn poor. However, it is impossible to deduce the Mn concentration in Ge$_{1-x}$Mn$_{x}$ nanocolumns from this cross section. Indeed, in cross section observations, the columns diameter is much smaller than the probed film thickness and the signal comes from the superposititon of the Ge matrix and Mn-rich nanocolumns. In order to quantify Mn concentration inside the nanocolumns and inside the Ge matrix, EELS measurements (not shown here) have been performed in a plane view geometry \cite{Jame06}. These observations revealed that the matrix Mn content is below 1 \% (detection limit of our instrument). Measuring the surface occupied by the matrix and the nanocolumns in plane view TEM images, and considering the average Mn concentration in the sample (6 \%), we can estimate the Mn concentration in the nanocolumns. The Mn concentration measured by EELS being between 0\% and 1\%, we can conclude that the Mn content in the nanocolumns is between 30 \% and 38 \%.\\
For samples grown between 80$^\circ$C and 150$^\circ$C cross section and plane view TEM observations reveal the presence of Mn rich nanocolumns surrounded with a Mn poor Ge matrix. In order to investigate the influence of Mn concentration on the structural properties of Ge$_{1-x}$Mn$_{x}$ films, ten samples have been grown at 100$^\circ$C and at 150$^\circ$C with Mn concentrations of 1.3 \%, 2.3 \%, 4 \%, 7 \% and 11.3 \%. Their structural properties have been investigated by plane view TEM observations. 

\begin{figure}[htb]
    \center
    \includegraphics[width=.98\linewidth]{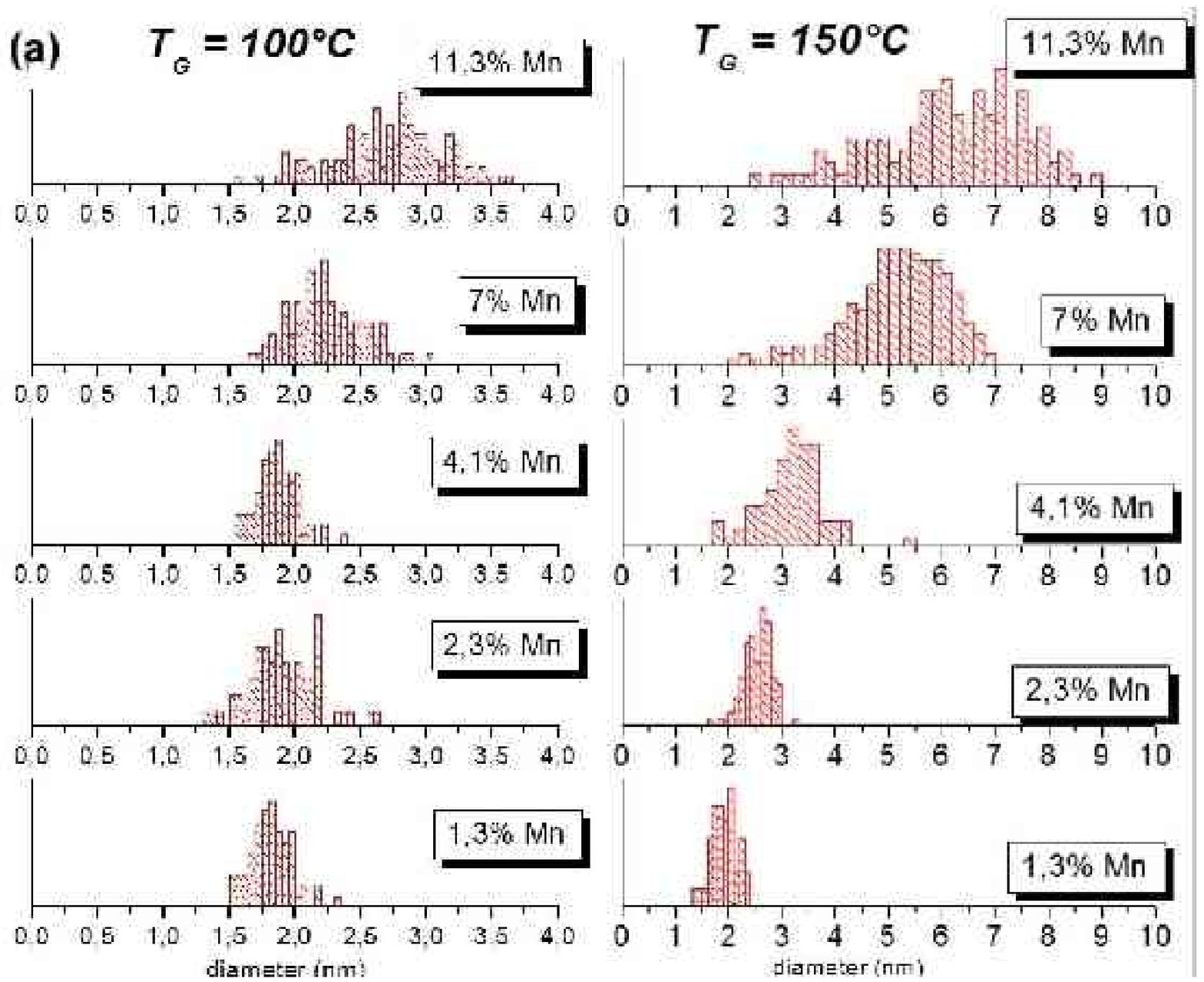}
	\includegraphics[width=.45\linewidth]{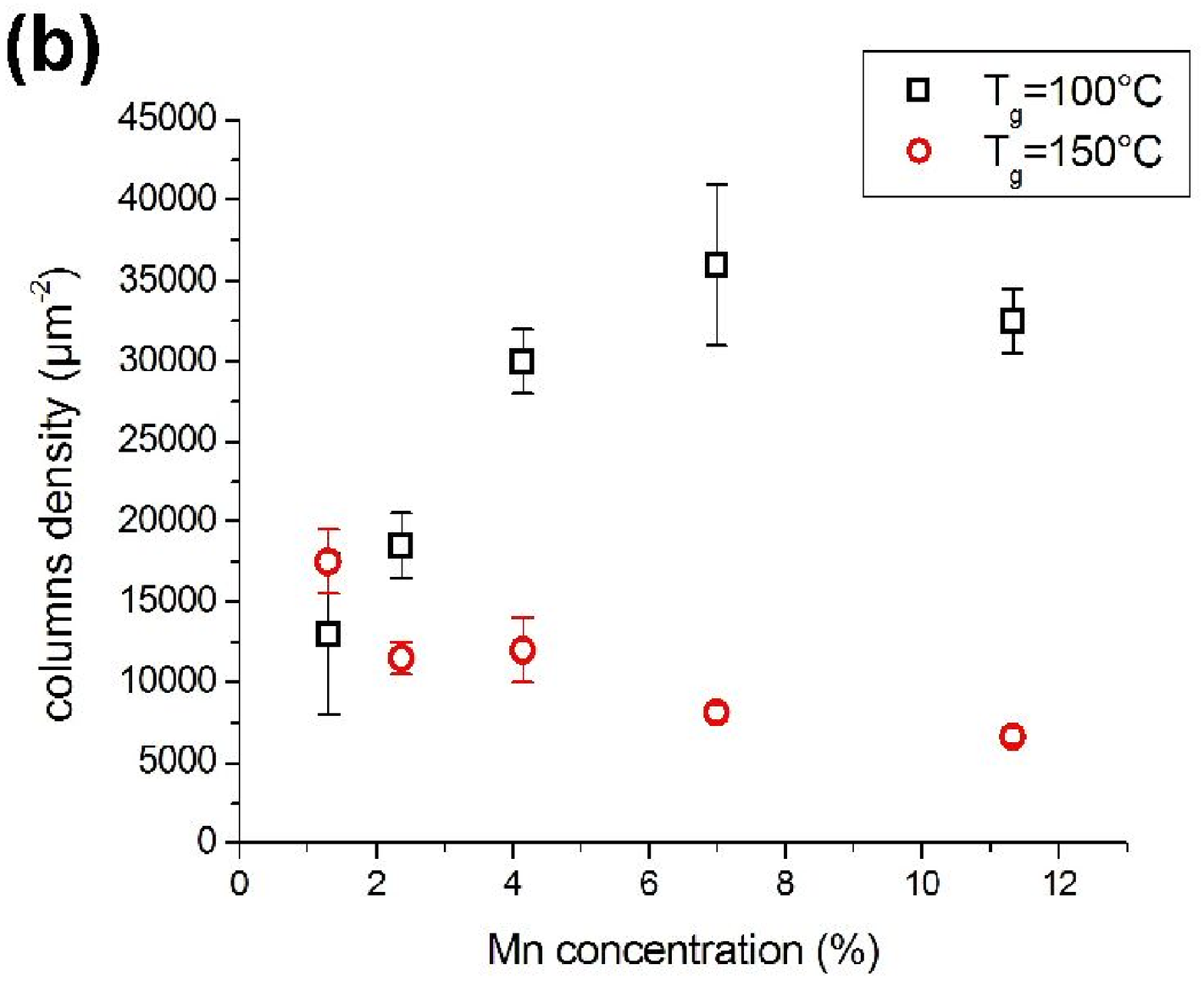}
		\includegraphics[width=.45\linewidth]{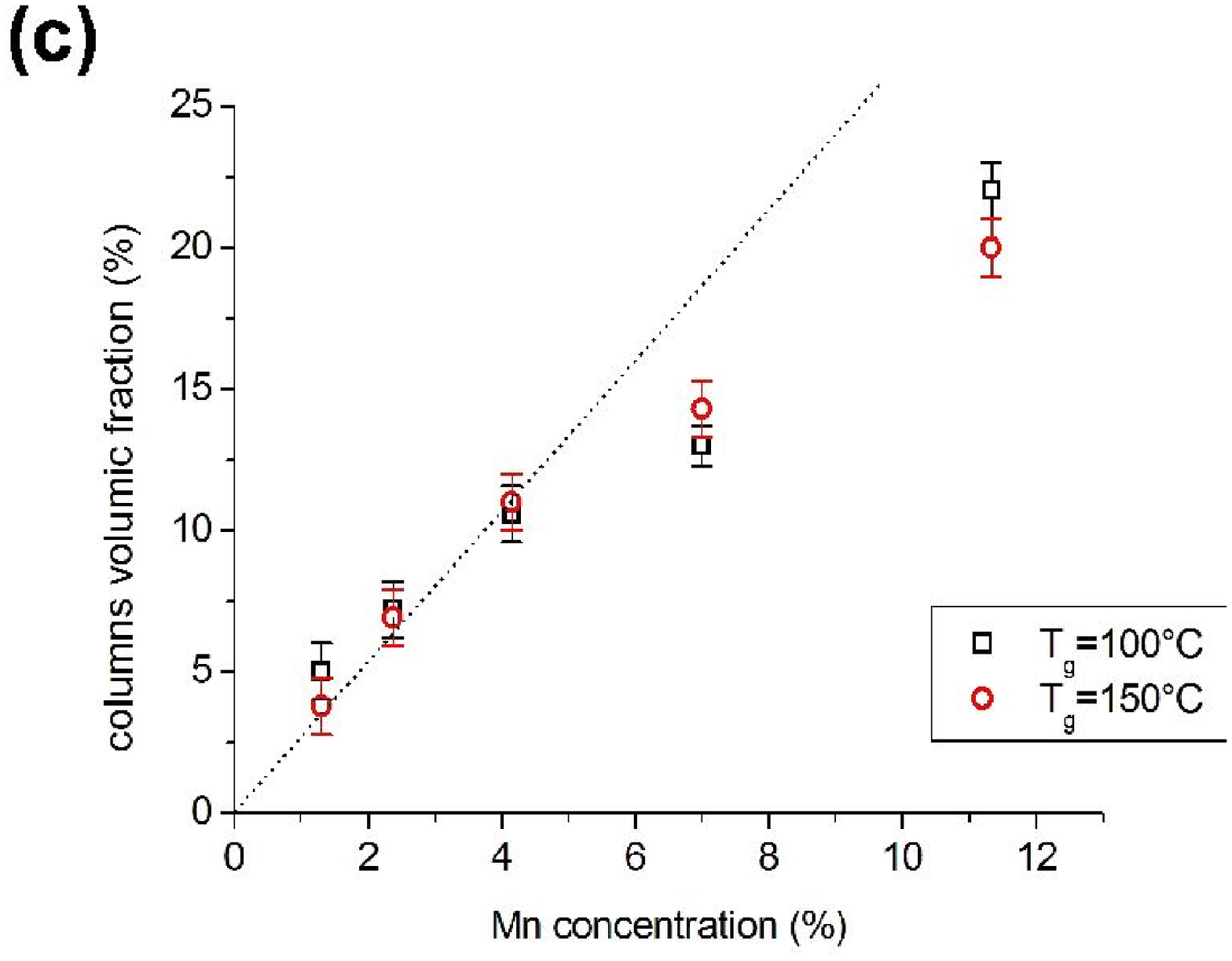}
    \caption{Nanocolumns size and density as a function of growth conditions. Samples considered have been grown at 100$^{\circ}$C and 150$^{\circ}$C respectively. (a) Mn concentration dependence of the size distribution. (b) columns density as a function of Mn concentration. (c) Volume fraction of the nanocolumns as a function of Mn concentration.}
 \label{fig3}
\end{figure}

For samples grown at 100$^\circ$C with Mn concentrations below 5 \% the nanocolumns mean diameter is 1.8$\pm$0.2 nm. The evolution of columns density as a fonction of Mn concentration is reported in figure 3b. By increasing the Mn concentration from 1.3 \% to 4 \% we observe a significant increase of the columns density from 13000 to 30000 $\rm{\mu m}^{-2}$. For Mn concentrations higher than 5 \% the density seems to reach a plateau corresponding to 35000 $\rm{\mu m}^{-2}$ and their diameter slightly increases from 1.8 nm at 4 \% to 2.8 nm at 11.3 \%. By plotting the volume fraction occupied by the columns in the film as a function of Mn concentration, we observe a linear dependence for Mn contents below 5 \%. The non-linear behavior above 5 \% may indicate that the mechanism of Mn incorporation is different in this concentration range, leading to an increase of Mn concentration in the columns or in the matrix. For samples grown at 100$^\circ$C, nanocolumns are always fully coherent with the surrounding matrix (Fig. 4a). 

Increasing the Mn content in the samples grown at 150$^\circ$C from 1.3 \% to 11.3 \% leads to a decrease of the columns density (fig 3b). Moreover, their average diameter increases significantly and size distributions become very broad (see Fig. 3a). For the highest Mn concentration (11.3 \%) we observe the coexistence of very small columns with a diameter of 2.5 nm and very large columns with a diameter of 9 nm. In samples grown at 150$^\circ$C containing 11.3 \% of Mn, the crystalline structure of nanocolumns is also highly modified. In plane view TEM micrographs, one can see columns exhibiting several different crystalline structures. We still observe some columns which are fully coherent with the Ge matrix like in the samples grown at lower temperature. Nevertheless, observations performed on these samples grown at 150$^\circ$C and with 11.3\% Mn reveal some uniaxially \cite{Jame06} or fully relaxed columns exhibiting a misfit of 4 \% between the matrix and the columns and leading to misfit dislocations at the interface between the column and the matrix (see fig. 4b). Thus we can conclude that coherent columns are probably in strong compression and the surrounding matrix in tension. On the same samples (T$_g$=150$^{\circ}$C, 11.3\% Mn), we also observe a large number of highly disordered nanocolumns leading to an amorphous like TEM contrast(fig. 4c).

\begin{figure}[htb]
    \center
   \includegraphics[width=.31\linewidth]{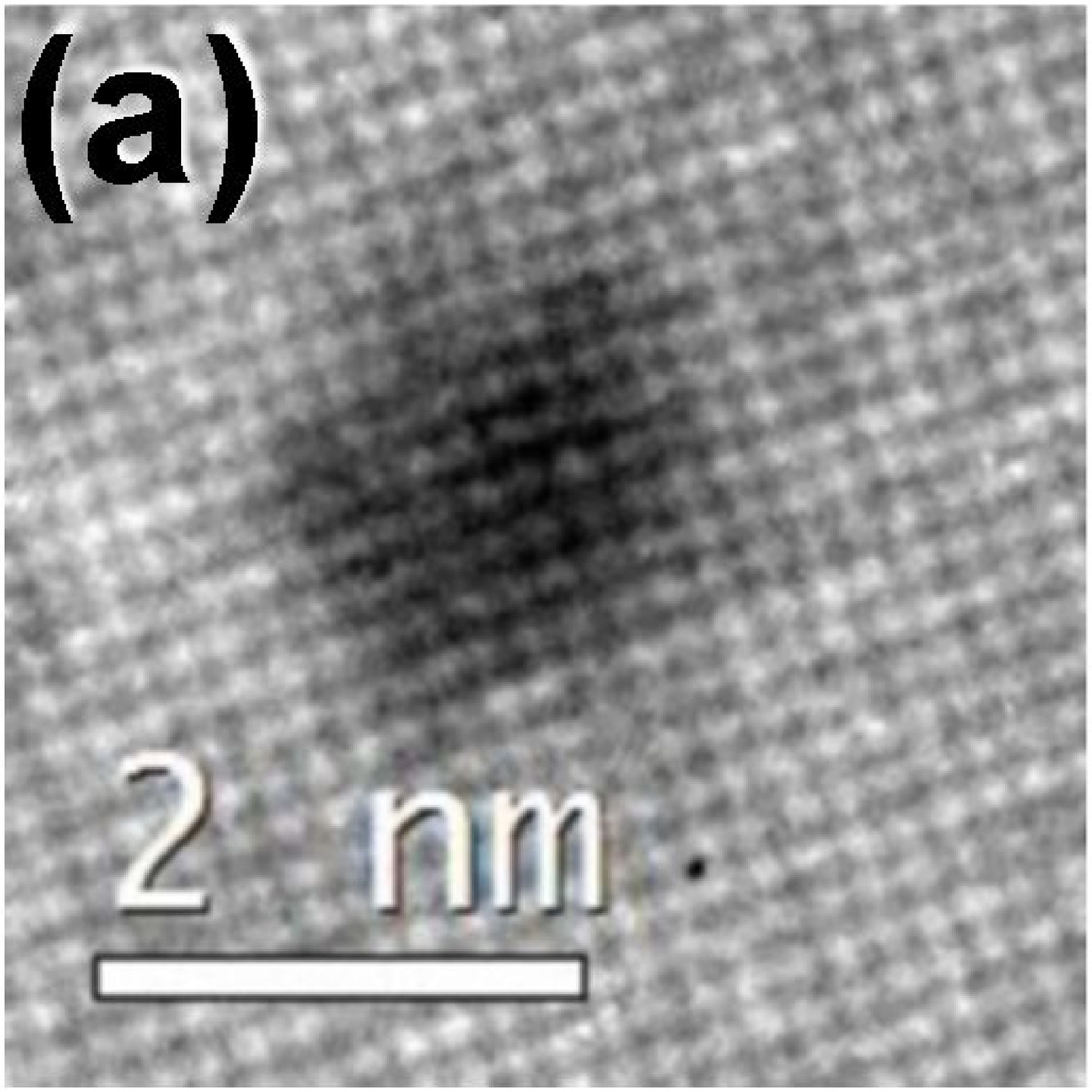}
	\includegraphics[width=.31\linewidth]{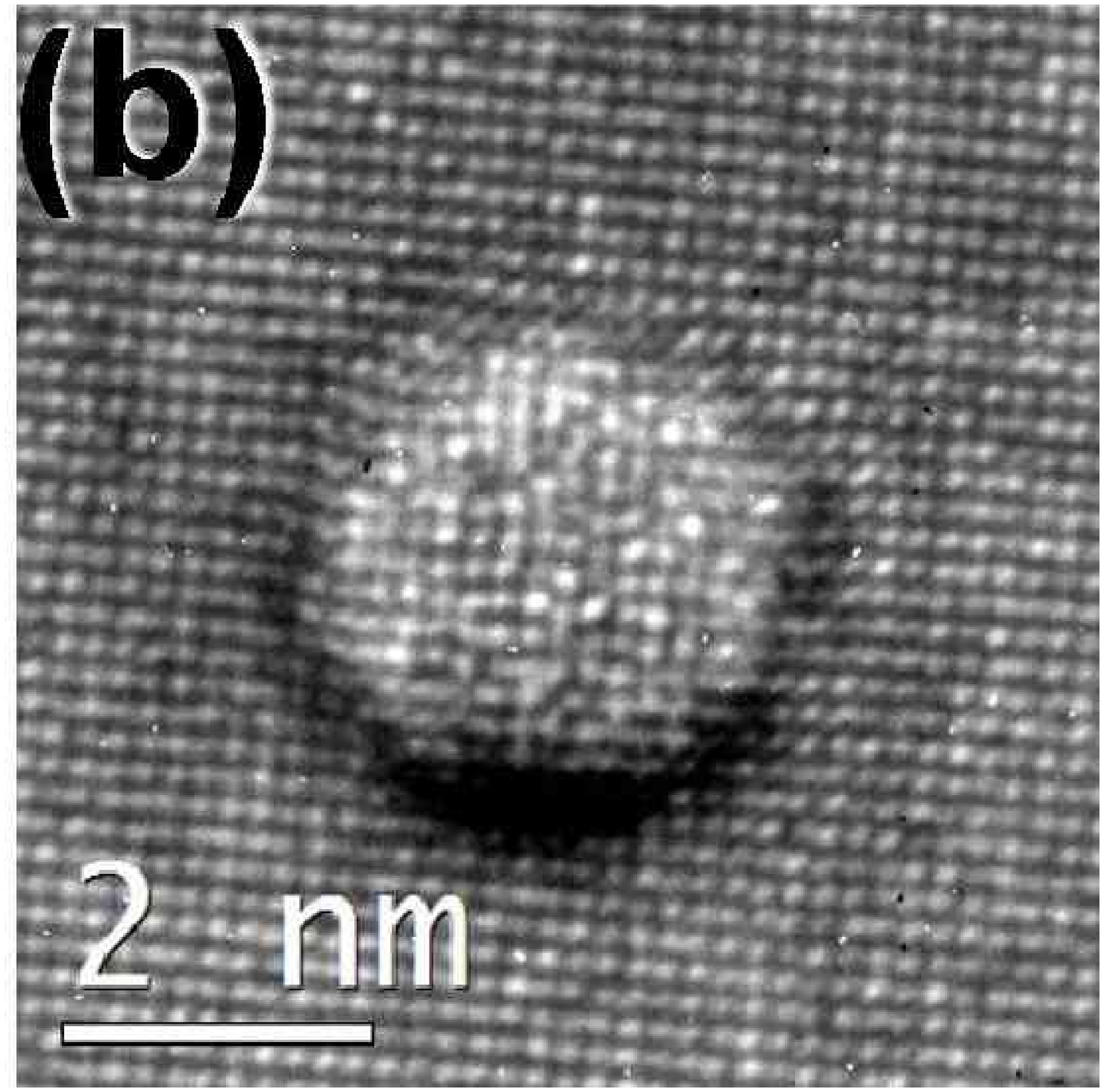}
	\includegraphics[width=.31\linewidth]{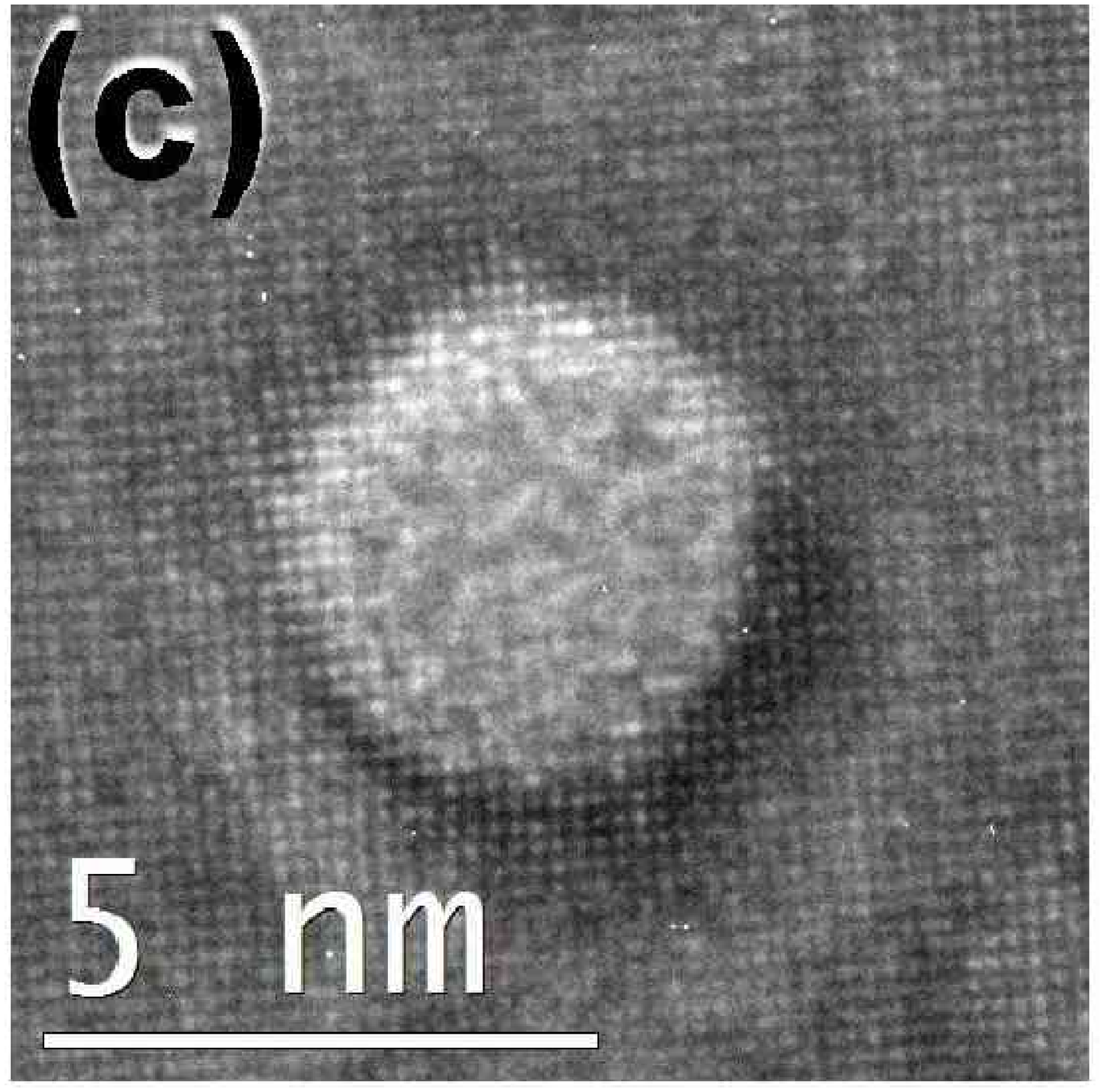}
    \caption{Plane view high resolution transmission electron micrographs of different types of nanocolumns : (a) typical structure of a column grown at 100$^{\circ}$C. The crystal structure is exactly the same as germanium . (b) Partially relaxed nanocolumn. One can see dislocations at the interface between the columns and the matrix leading to stress relaxation. (c) Amorphous nanocolumn. These columns are typical in samples grown at 150$^{\circ}$C with high Mn contents.}
 \label{fig4}
\end{figure}

In conclusion, we have evidenced a complex mechanism of Mn incorporation in Mn doped Ge films grown at low temperature. In particular Mn incorporation is highly inhomogeneous. For very low growth temperatures (below 120$^\circ$C) the diffusion of Mn atoms leads to the formation of Mn rich, vertical nanocolumns. Their density mostly depends on Mn concentration and their mean diameter is about 2 nm. These results can be compared with the theoretical predictions of Fukushima \textit{et al.} \cite{Fuku06}: they proposed a model of spinodal decomposition in (Ga,Mn)N and (Zn,Cr)Te based on layer by layer growth conditions and a strong pair attraction between Mn atoms which leads to the formation of nanocolumns. This model may also properly describe the formation of Mn rich nanocolumns in our samples. Layer by layer growth conditions can be deduced from RHEED pattern evolution during growth. For all the samples grown at low temperature, RHEED observations clearly indicate two-dimensional growth. Moreover, Ge/Ge$_{1-x}$Mn$_{x}$/Ge heterostructures have been grown and observed by TEM (see Fig. 5). Ge$_{1-x}$Mn$_{x}$/Ge (as well as Ge/Ge$_{1-x}$Mn$_{x}$) interfaces are very flat and sharp thus confirming a two-dimensional, layer by layer growth mode. Therefore we can assume that the formation of Mn rich nanocolumns is a consequence of 2D-spinodal decomposition.

\begin{figure}[htb]
    \center
	\includegraphics[width=.7\linewidth]{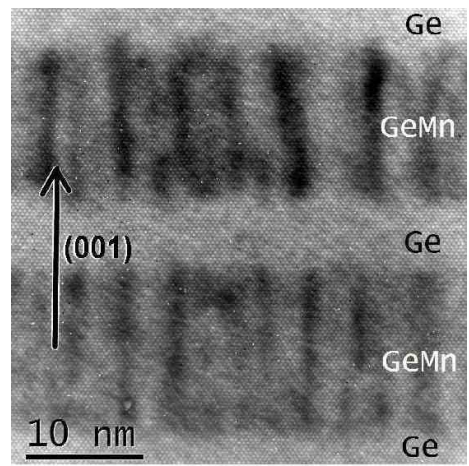}
    \caption{Cross section high resolution micrograph of a Ge/Ge$_{1-x}$Mn$_{x}$/Ge/Ge$_{1-x}$Mn$_{x}$/Ge heterostructure. This sample has been grown at 130 $^{\circ}$C with 6\% Mn. Ge$_{1-x}$Mn$_{x}$ layers are 15 nm thick and Ge spacers 5 nm thick. We clearly see the sharpness of both Ge$_{1-x}$Mn$_{x}$/Ge and Ge/Ge$_{1-x}$Mn$_{x}$ interfaces. Mn segregation leading to the columns formation already takes place in very thin Ge$_{1-x}$Mn$_{x}$ films.}
\label{fig5}
\end{figure}

For growth temperatures higher than 160$^\circ$C, cross section TEM and EFTEM observations (not shown here) reveal the coexistence of two Mn-rich phases: nanocolumns and Ge$_{3}$Mn$_{5}$ nanoclusters embedded in the germanium matrix. A typical high resolution TEM image is shown in figure 6. 
Ge$_{3}$Mn$_{5}$ clusters are not visible in RHEED patterns for temperatures below 180$^\circ$C. To investigate the nature of these clusters, we performed x-ray diffraction in $\theta-2\theta$ mode. Diffraction scans were acquired on a high resolution diffractometer using the copper K$_\alpha$ radiation and on the GMT station of the BM32 beamline at the European Synchrotron Radiation Facility (ESRF). Three samples grown at different temperatures and/or annealed at high temperature were investigated. The two first samples are Ge$_{1-x}$Mn$_{x}$ films grown at 130$^\circ$C and 170$^\circ$C respectively. The third one has been grown at 130$^\circ$C and post-growth annealed at 650$^\circ$C. By analysing x-ray diffraction spectra, we can evidence two different crystalline structures. For the sample grown at 130$^\circ$C, the $\theta-2\theta$ scan only reveals the (004) Bragg peak of the germanium crystal, confirming the good epitaxial relationship between the layer and the substrate, and the absence of secondary phases in the film in spite of a high dynamics of the order of 10$^7$. For both samples grown at 170$^\circ$C and annealed at 650$^\circ$C, $\theta-2\theta$ spectra are identical. In addition to the (004) peak of germanium, we observe three additional weak peaks. The first one corresponds to the (002) germanium forbidden peak which probably comes from a small distortion of the germanium crystal, and the two other peaks are respectively attributed to the (002) and (004) Bragg peaks of a secondary phase. The $c$ lattice parameter of Ge$_3$Mn$_5$ hexagonal crystal is 5.053 \AA \ \cite{Fort90} which is in very good agreement with the values obtained from diffraction data for both (002) and (004) lines assuming that the $c$ axis of Ge$_3$Mn$_5$ is along the [001] direction of the Ge substrate.

\begin{figure}[htb]
    \center
	\includegraphics[width=.7\linewidth]{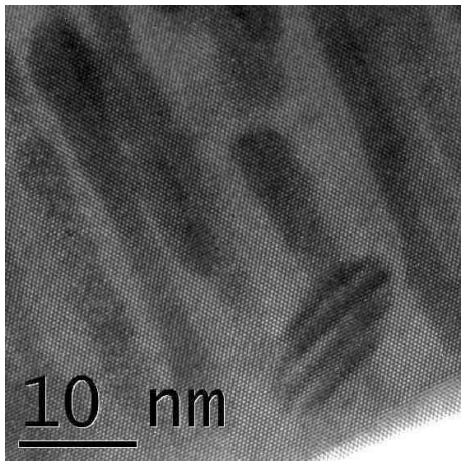}
	\caption{Cross section high resolution transmission electron micrograph of a sample grown at 170$^{\circ}$C. We observe the coexistence of two different Mn-rich phases: Ge$_{1-x}$Mn$_{x}$ nanocolumns and Ge$_3$Mn$_5$ clusters.}
\label{fig6}
\end{figure}

In summary, in a wide range of growth temperatures and Mn concentrations, we have evidenced a two-dimensional spinodal decomposition leading to the formation of Mn-rich nanocolumns in Ge$_{1-x}$Mn$_{x}$ films. This decomposition is probably the consequence of: $(i)$ a strong pair attraction between Mn atoms, $(ii)$ a strong surface diffusion of Mn atoms in germanium even at low growth temperatures and $(iii)$ layer by layer growth conditions. We have also investigated the influence of growth parameters on the spinodal decomposition: at low growth temperatures (100$^{\circ}$C), increasing the Mn content leads to higher columns densities while at higher growth temperatures (150$^{\circ}$C), the columns density remains nearly constant whereas their size increases drastically. By plotting the nanocolumns density as a function of Mn content, we have shown that the mechanism of Mn incorporation in Ge changes above 5 \% of Mn. Finally, using TEM observations and x-ray diffraction, we have shown that Ge$_3$Mn$_5$ nanoclusters start to form at growth temperatures higher than 160$^\circ$C.

\section{Magnetic properties \label{magnetic}}

We have thoroughly investigated the magnetic properties of thin Ge$_{1-x}$Mn$_{x}$ films for different growth temperatures and Mn concentrations. In this section, we focus on Mn concentrations between 2 \% and 11 \%. We could clearly identify four different magnetic phases in Ge$_{1-x}$Mn$_{x}$ films : diluted Mn atoms in the germanium matrix, low $T_{C}$ nanocolumns ($T_{C}$ $\leq$ 170 K), high $T_{C}$ nanocolumns ($T_{C}$ $\geq$ 400 K) and Ge$_{3}$Mn$_{5}$ clusters ($T_{C}$ $\thickapprox$ 300 K). The relative weight of each phase clearly depends on the growth temperature and to a lesser extend on Mn concentration. For low growth temperature ($<$ 120$^{\circ}$C), we show that nanocolumns are actually made of four uncorrelated superparamagnetic nanostructures. Increasing T$_{g}$ above 120$^{\circ}$C, we first obtain continuous columns exhibiting low $T_{C}$ ($<$ 170 K) and high $T_{C}$ ($>$ 400 K) for $T_{g}\approx$130$^{\circ}$C. The larger columns become ferromagnetic \textit{i.e.} $T_{B}>T_{C}$. Meanwhile Ge$_{3}$Mn$_{5}$ clusters start to form. Finally for higher $T_{g}$, the magnetic contribution from Ge$_{3}$Mn$_{5}$ clusters keeps increasing while the nanocolumns signal progressively disappears.

\begin{figure}[htb]
\center
   \includegraphics[width=.6\linewidth]{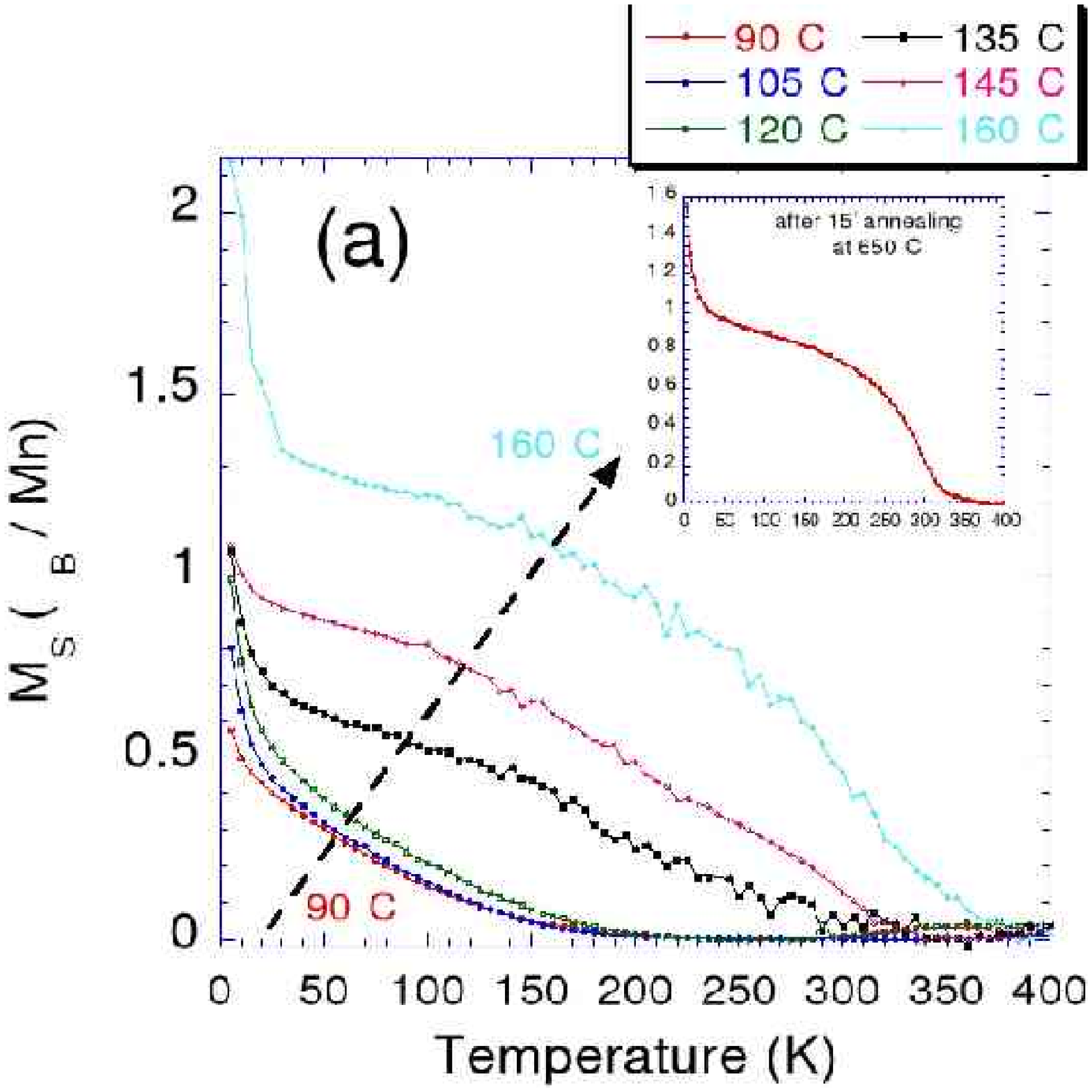}
   \includegraphics[width=.3\linewidth]{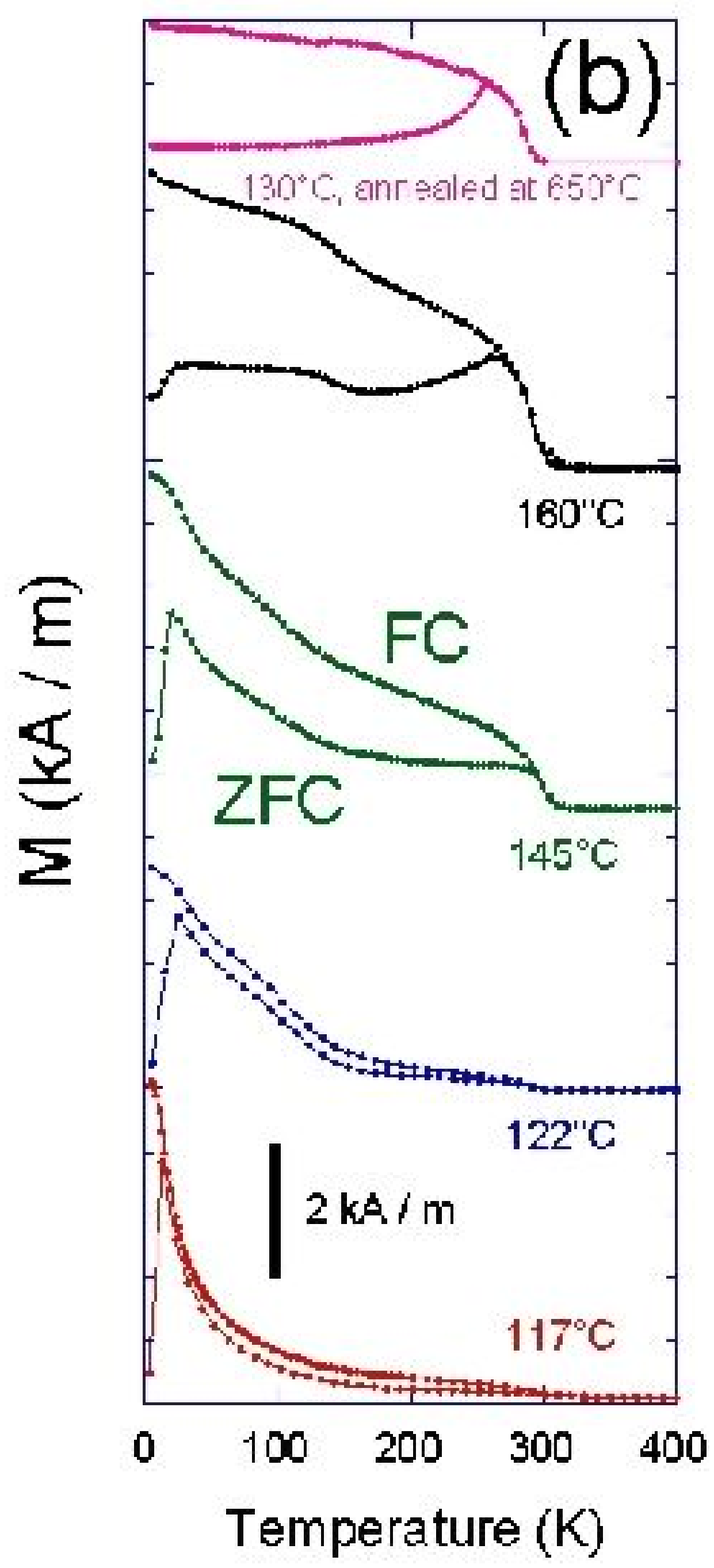}
\caption{(a) Temperature dependence of the saturation magnetization (in $\mu_{B}$/Mn) of Ge$_{0.93}$Mn$_{0.07}$ samples for different growth temperatures. The magnetic field is applied in the film plane. The inset shows the temperature dependence of a sample grown at 130$^{\circ}$C and annealed at 650$^{\circ}$C for 15 minutes. After annealing, the magnetic signal mostly arises from Ge$_{3}$Mn$_{5}$ clusters. (b) ZFC-FC measurements performed on Ge$_{0.93}$Mn$_{0.07}$ samples for different growth temperatures. The in-plane applied field is 0.015 T. The ZFC peak at low temperature ($\leq$150 K) can be attributed to the superparamagnetic nanocolumns. This peak widens and shifts towards high blocking temperatures when increasing growth temperature. The second peak above 150 K in the ZFC curve which increases with increasing growth temperature is attributed to superparamagnetic Ge$_{3}$Mn$_{5}$ clusters. The increasing ZFC-FC irreversibility at $\approx$ 300 K is due to the increasing contribution from large ferromagnetic Ge$_{3}$Mn$_{5}$ clusters. The nanocolumns signal completely vanishes after annealing at 650$^{\circ}$C for 15 minutes.}
\label{fig7}
\end{figure}

In Fig. 7a, the saturation magnetization at 2 Tesla in $\mu_{B}$/Mn of Ge$_{1-x}$Mn$_{x}$ films with 7 \% of Mn is plotted as a function of temperature for different growth temperatures ranging from $T_{g}$=90$^{\circ}$C up to 160$^{\circ}$C. The inset shows the temperature dependence of the magnetization at 2 Tesla after annealing at 650$^{\circ}$C during 15 minutes. Figure 7b displays the corresponding Zero Field Cooled - Field Cooled (ZFC-FC) curves recorded at 0.015 Tesla. In the ZFC-FC procedure, the sample is first cooled down to 5 K in zero magnetic field and the susceptibility is subsequently recorded at 0.015 Tesla while increasing the temperature up to 400 K (ZFC curve). Then, the susceptibility is recorded under the same magnetic field while decreasing the temperature down to 5 K (FC curve). Three different regimes can be clearly distinguished. \\
For $T_{g}\leq$120$^{\circ}$C, the temperature dependence of the saturation magnetization remains nearly the same while increasing growth temperature. The overall magnetic signal vanishing above 200 K is attributed to the nanocolumns whereas the increasing signal below 50 K originates from diluted Mn atoms in the surrounding matrix. The Mn concentration dependence of the saturation magnetization is displayed in figure 8. For the lowest Mn concentration (4 \%), the contribution from diluted Mn atoms is very high and drops sharply for higher Mn concentrations (7 \%, 9 \% and 11.3 \%). Therefore the fraction of Mn atoms in the diluted matrix decreases with Mn concentration probably because Mn atoms are more and more incorporated in the nanocolumns. In parallel, the Curie temperature of nanocolumns increases with the Mn concentration reaching 170 K for 11.3 \% of Mn. This behavior may be related to different Mn compositions and to the increasing diameter of nanocolumns (from 1.8 nm to 2.8 nm) as discussed in section \ref{structural}.

\begin{figure}[htb]
\center
   \includegraphics[width=.7\linewidth]{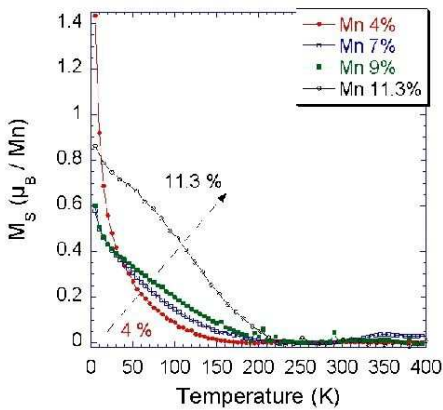}
    \caption{Temperature dependence of the saturation magnetization (in $\mu_{B}$/Mn) of Ge$_{1-x}$Mn$_{x}$ films grown at 100$^{\circ}$C plotted for different Mn concentrations: 4.1 \%; 7 \%; 8.9 \% and 11.3 \%.}
\label{fig8}
\end{figure}

ZFC-FC measurements show that the nanocolumns are superparamagnetic. The magnetic signal from the diluted Mn atoms in the matrix is too weak to be detected in susceptibility measurements at low temperature. In samples containing 4 \% of Mn, ZFC and FC curves superimpose down to low temperatures. As we do not observe hysteresis loops at low temperature, we believe that at this Mn concentration nanocolumns are  superparamagnetic in the whole temperature range and the blocking temperature cannot be measured. For higher Mn contents, the ZFC curve exhibits a very narrow peak with a maximum at the blocking temperature of 15 K whatever the Mn concentration and growth temperature (see Fig. 7b). Therefore the anisotropy barrier distribution is narrow and assuming that nanocolumns have the same magnetic anisotropy, this is a consequence of the very narrow size distribution of the nanocolumns as observed by TEM. To probe the anisotropy barrier distribution, we have performed ZFC-FC measurements but instead of warming the sample up to 400 K, we stopped at a lower temperature $T_{0}$. 

\begin{figure}[htb]
\center
   \includegraphics[width=.6\linewidth]{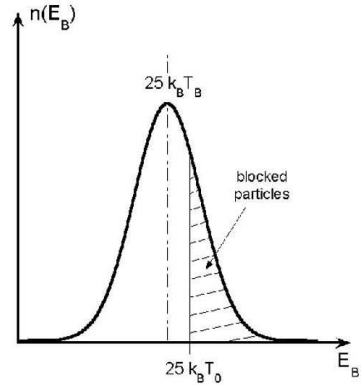}
\caption{Schematic drawing of the anisotropy barrier distribution n($E_{B}$) of superparamagnetic nanostructures. If magnetic anisotropy does not depend on the particle size, this distribution exactly reflects their magnetic size distribution. In this drawing the blocking temperature ($T_{B}$) corresponds to the distribution maximum. At a given temperature $T_{0}$ such that 25$k_{B}T_{0}$ falls into the anisotropy barrier distribution, the largest nanostructures with an anisotropy energy larger than 25$k_{B}T_{0}$ are blocked whereas the others are superparamagnetic.}
\label{fig9}
\end{figure}

If this temperature falls into the anisotropy barrier distribution as depicted in Fig. 9, the FC curve deviates from the ZFC curve. Indeed the smallest nanostructures have become superparamagnetic at $T_{0}$ and when decreasing again the temperature, their magnetization freezes along a direction close to the magnetic field and the FC susceptibility is higher than the ZFC susceptibility. Therefore any irreversibility in this procedure points at the presence of superparamagnetic nanostructures. The results are given in Fig. 10a. ZFC and FC curves clearly superimpose up to $T_{0}$=250 K thus the nanocolumns are superparamagnetic up to their Curie temperature and no Ge$_{3}$Mn$_{5}$ clusters could be detected. Moreover for low $T_{0}$ values, a peak appears at low temperature in FC curves which evidences strong antiferromagnetic interactions between the nanocolumns \cite{Chan00}.

\begin{figure}[htb]
\center
    \includegraphics[width=.35\linewidth]{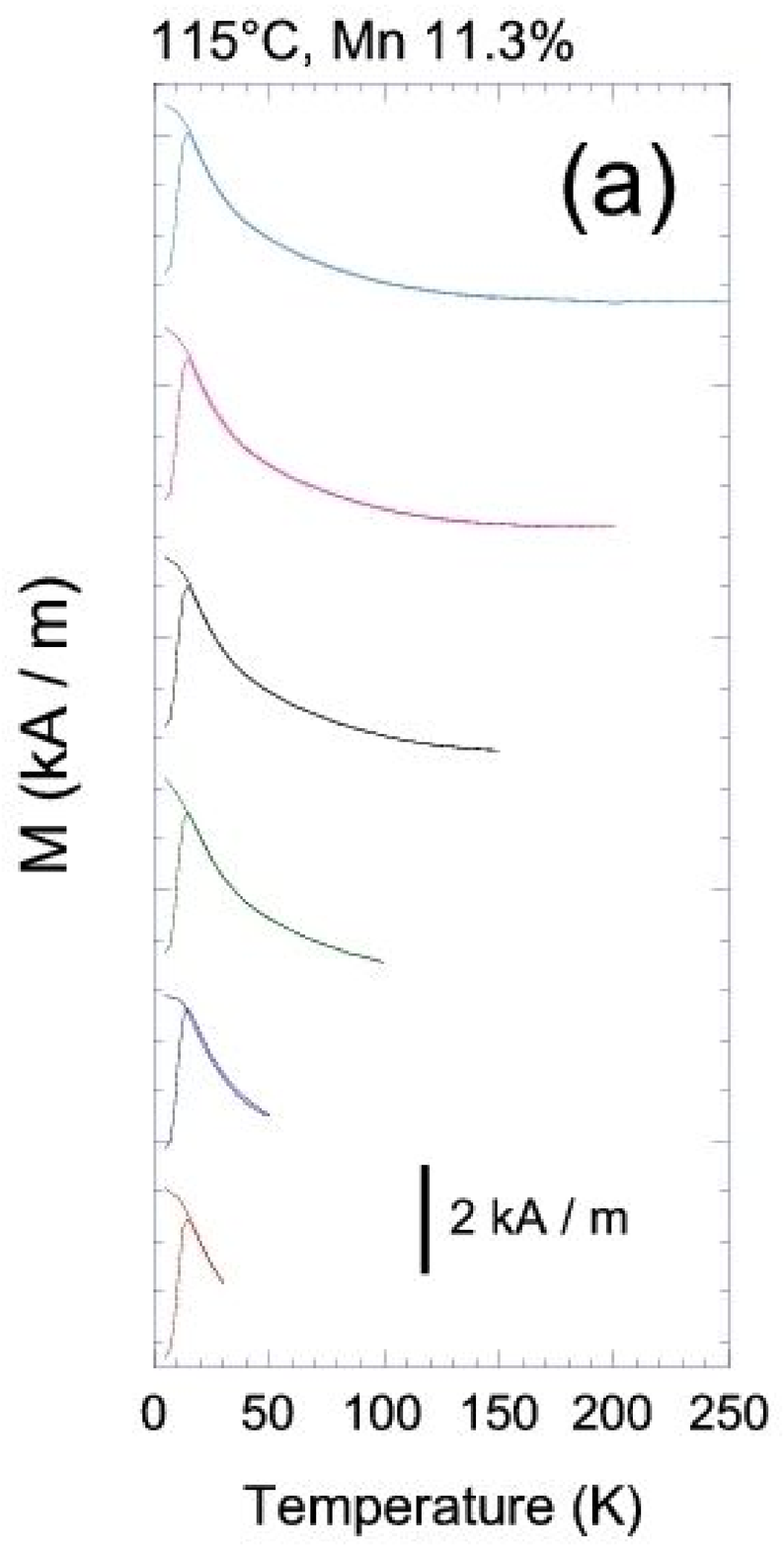}
    \includegraphics[width=.63\linewidth]{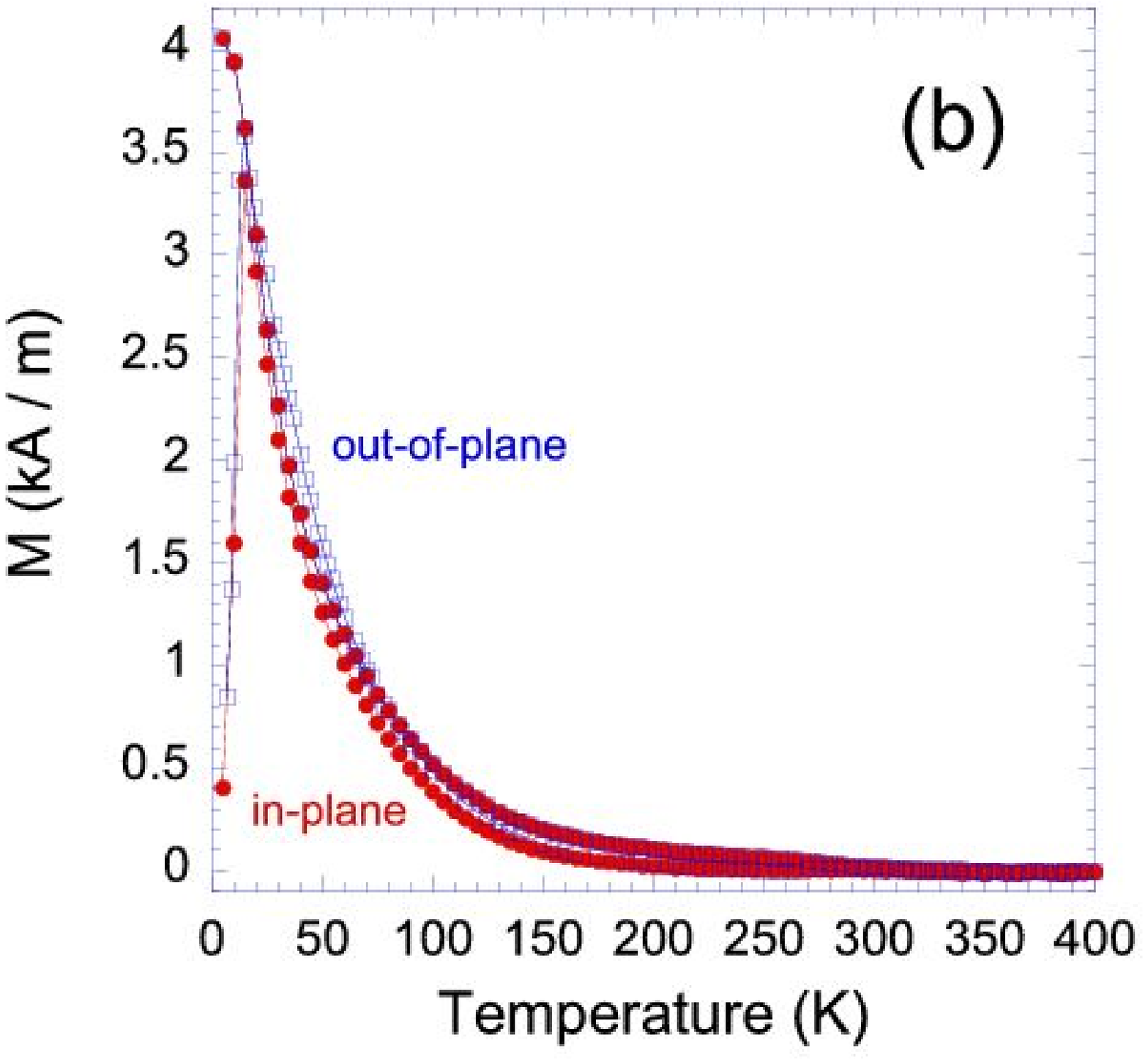}
\caption{(a) ZFC-FC measurements performed on a Ge$_{0.887}$Mn$_{0.113}$ sample grown at 115$^{\circ}$C. The in-plane applied field is 0.015 T. Magnetization was recorded up to different T$_{0}$ temperatures: 30 K, 50 K, 100 K, 150 K, 200 K and 250 K. Curves are shifted up for more clarity. (b) ZFC-FC curves for in-plane and out-of-plane applied fields (0.015 T).}
\label{fig10}
\end{figure}

In order to derive the magnetic size and anisotropy of the Mn-rich nanocolumns embedded in the Ge matrix, we have fitted the inverse normalized in-plane (resp. out-of-plane) susceptibility: $\chi_{\parallel}^{-1}$ (resp. $\chi_{\perp}^{-1}$). The corresponding experimental ZFC-FC curves are reported in Fig. 10b. Since susceptibility measurements are performed at low field (0.015 T), the matrix magnetic signal remains negligible. In order to normalize susceptibility data, we need to divide the magnetic moment by the saturated magnetic moment recorded at 5 T. However the matrix magnetic signal becomes very strong at 5 T and low temperature so that we need to subtract it from the saturated magnetic moment using a simple Curie function. From Fig. 10b, we can conclude that nanocolumns are isotropic. Therefore to fit experimental data we use the following expression well suited for isotropic systems or cubic anisotropy: $\chi_{\parallel}^{-1}= \chi_{\perp}^{-1}\approx 3k_{B}T/M(T)+\mu_{0}H_{eff}(T)$. $k_{B}$ is the Boltzmann constant, $M=M_{s}v$ is the magnetic moment of a single-domain nanostructure (macrospin approximation) where $M_{s}$ is its magnetization and $v$ its volume. The in-plane magnetic field is applied along $[110]$ or $[-110]$ crystal axes. Since the nanostructures Curie temperature does not exceed 170 K, the temperature dependence of the saturation magnetization is also accounted for by writting $M(T)$. Antiferromagnetic interactions between nanostructures are also considered by adding an effective field estimated in the mean field approximation \cite{Fruc02}: $\mu_{0}H_{eff}(T)$.
The only fitting parameters are the maximum magnetic moment (\textit{i.e.} at low temperature) per nanostructure: $M$ (in Bohr magnetons $\mu_{B}$) and the maximum interaction field (\textit{i.e.} at low temperature): $\mu_{0}H_{eff}$.

\begin{figure}[htb]
\center
   \includegraphics[width=.7\linewidth]{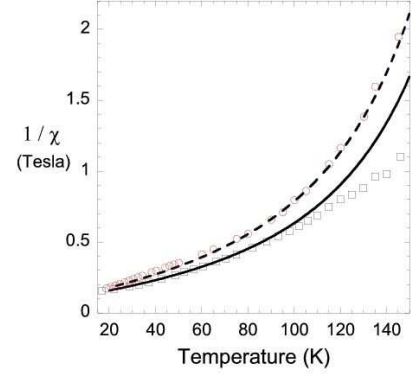}
\caption{Temperature dependence of the inverse in-plane (open circles) and out-of-plane (open squares) normalized susceptibilities of a Ge$_{0.887}$Mn$_{0.113}$ sample grown at 115$^{\circ}$C. Fits were performed assuming isotropic nanostructures or cubic anisotropy. Dashed line is for in-plane susceptibility and solid line for out-of-plane susceptibility.}
\label{fig11}
\end{figure}

In Fig. 11, the best fits lead to $M\approx$1250 $\mu_{B}$ and $\mu_{0}H_{eff}\approx$102 mT for in-plane susceptibility and $M\approx$1600 $\mu_{B}$ and $\mu_{0}H_{eff}\approx$98 mT for out-of-plane susceptibility. It gives an average magnetic moment of 1425 $\mu_{B}$ per column and an effective interaction field of 100 mT. Using this magnetic moment and its temperature dependence, magnetization curves could be fitted using a Langevin function and $M(H/T)$ curves superimpose for $T<$100 K. However, from the saturated magnetic moment of the columns and their density (35000 $\rm{\mu m}^{-2}$), we find almost 6000 $\mu_{B}$ per column. Therefore, for low growth temperatures, we need to assume that nanocolumns are actually made of almost four independent elongated magnetic nanostructures. The effective field for antiferromagnetic interactions between nanostructures estimated from the susceptibility fits is at least one order of magnitude larger than what is expected from pure magnetostatic coupling. This difference may be due to either an additional antiferromagnetic coupling through the matrix which origin remains unexplained or to the mean field approximation which is no more valid in this strong coupling regime. As for magnetic anisotropy, the nanostructures behave as isotropic magnetic systems or exhibit a cubic magnetic anisotropy. First we can confirm that nanostructures are not amorphous otherwise shape anisotropy would dominate leading to out-of-plane anisotropy. We can also rule out a random distribution of magnetic easy axes since the nanostructures are clearly crystallized in the diamond structure and would exhibit at least a cubic anisotropy (except if the random distribution of Mn atoms within the nanostructures can yield random easy axes). Since the nanostructures are in strong in-plane compression (their lattice parameter is larger than the matrix one), the cubic symmetry of the diamond structure is broken and magnetic cubic anisotropy is thus unlikely. We rather believe that out-of-plane shape anisotropy is nearly compensated by in-plane magnetoelastic anisotropy due to compression leading to a \textit{pseudo} cubic anisotropy. From the blocking temperature (15 K) and the magnetic volume of the nanostructures , we can derive their magnetic anisotropy constant using $Kv=25k_{B}T_{B}$: K$\approx$10 kJ.m$^{-3}$ which is of the same order of magnitude as shape anisotropy.

\begin{figure}[htb]
\center
    \includegraphics[width=.35\linewidth]{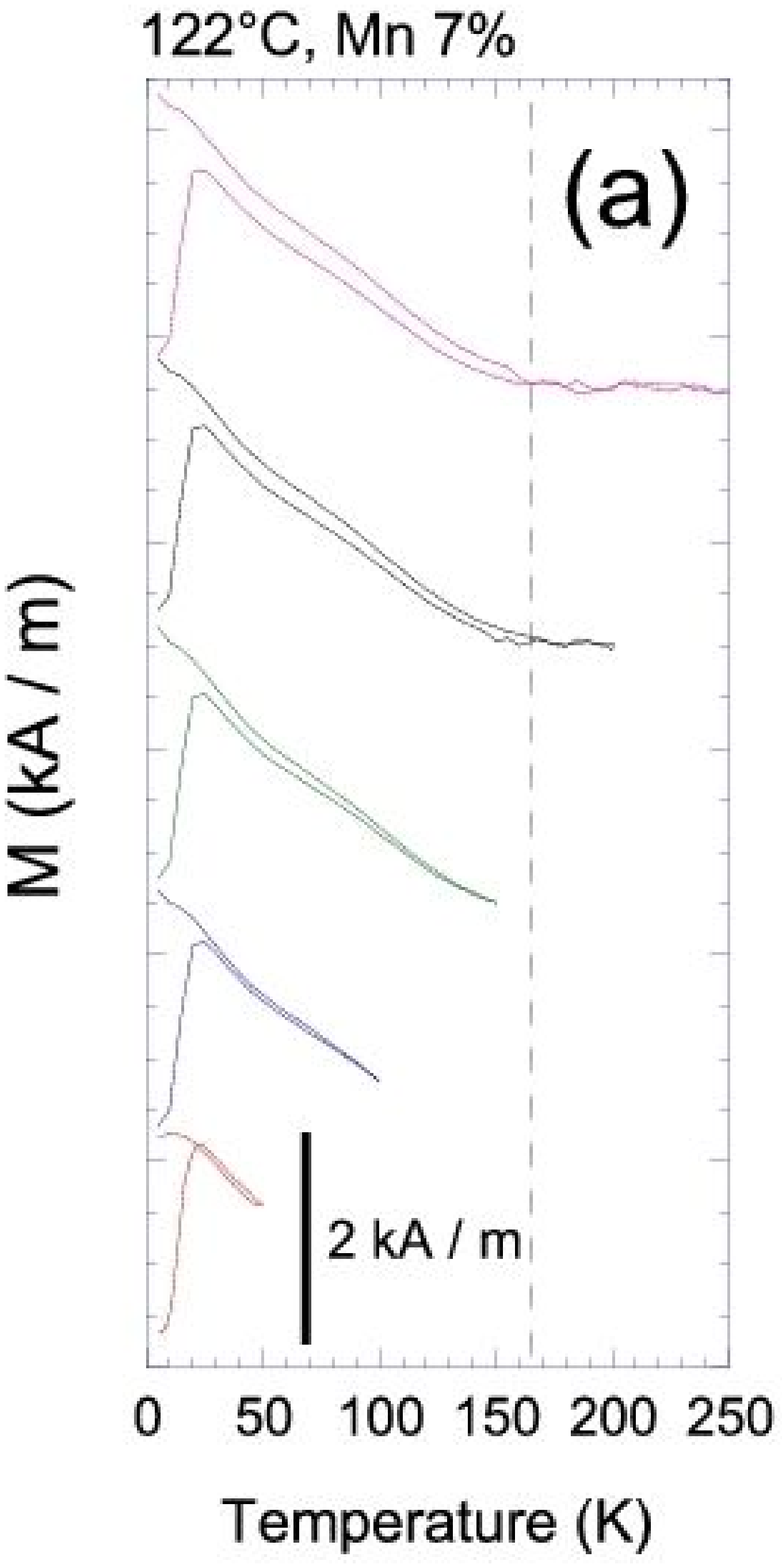}
    \includegraphics[width=.63\linewidth]{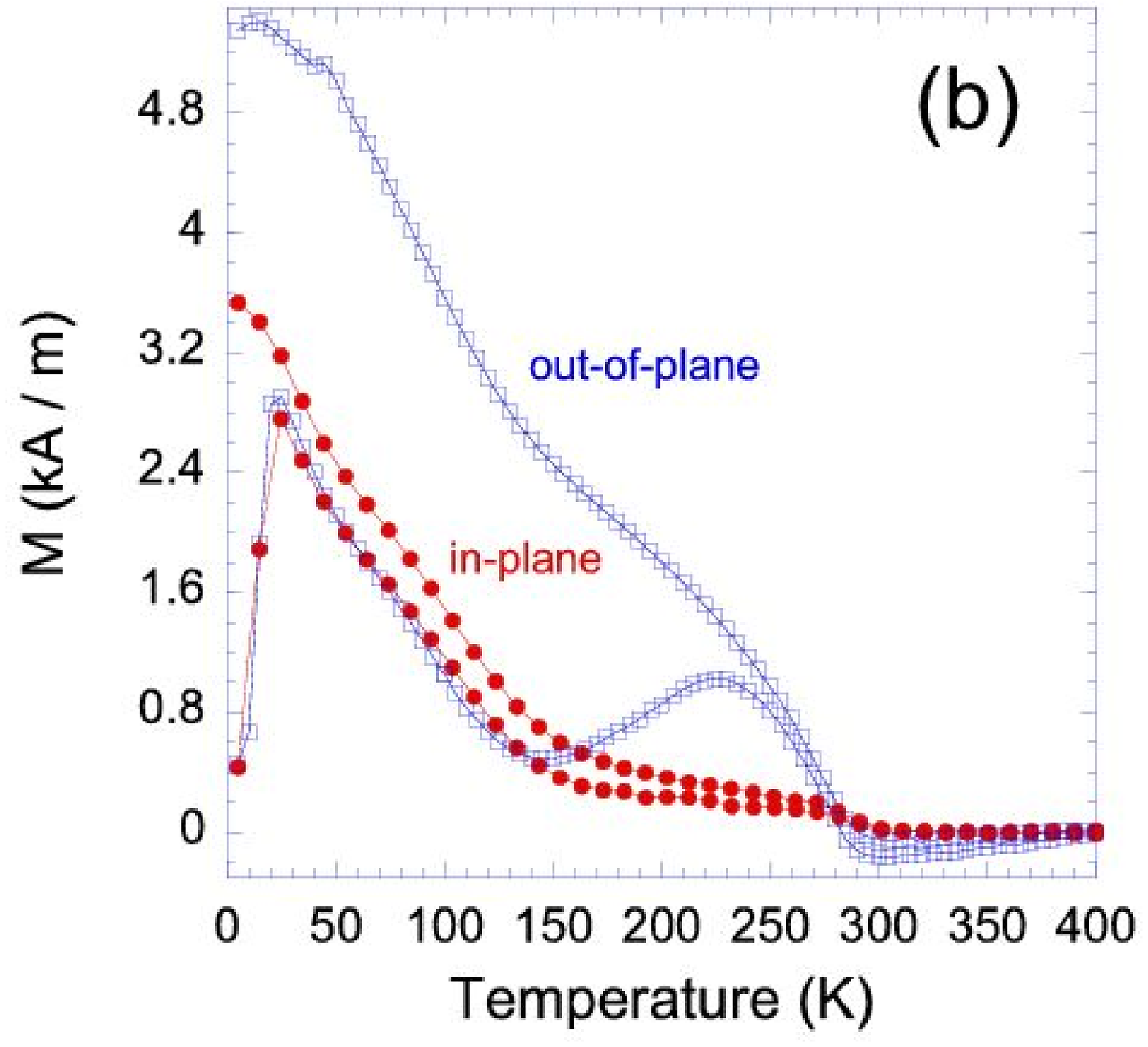} 
\caption{(a) ZFC-FC measurements performed on a Ge$_{0.93}$Mn$_{0.07}$ sample grown at 122$^{\circ}$C. The in-plane applied field is 0.015 T. Magnetization was recorded up to different T$_{0}$ temperatures: 50 K, 100 K, 150 K, 200 K and 250 K. Curves are shifted up for more clarity. (b) ZFC-FC curves for in-plane and out-of-plane applied fields  (0.015 T).}
\label{fig12}
\end{figure}

For growth temperatures $T_{g}\geq$120$^{\circ}$C and Mn concentrations $\geq$ 7 \%, samples exhibit a magnetic signal above 200 K corresponding to Ge$_{3}$Mn$_{5}$ clusters (see Fig. 7a). As we can see, SQUID measurements are much more sensitive to the presence of Ge$_{3}$Mn$_{5}$ clusters, even at low concentration, than TEM and x-ray diffraction used in section \ref{structural}. We also observe a sharp transition in the ZFC curve (see Fig. 7b, Fig. 12a and 12b): the peak becomes very large and is shifted towards high blocking temperatures (the signal is maximum at $T=$23 K). This can be easily understood as a magnetic percolation of the four independent nanostructures obtained at low growth temperatures into a single magnetic nanocolumn. Therefore the magnetic volume increases sharply as well as blocking temperatures. At the same time, the size distribution widens as observed in TEM. In Fig. 12a, we have performed ZFC-FC measurements at different $T_{0}$ temperatures. The ZFC-FC irreversibility is observed up to the Curie temperature of $\approx$120 K meaning that a fraction of nanocolumns is ferromagnetic (\textit{i.e.} $T_{B}\geq T_{C}$).
In Fig. 12b, in-plane and out-of-plane ZFC curves nearly superimpose for $T\leq$150 K due to the isotropic magnetic behavior of the nanocolumns: in-plane magnetoelastic anisotropy is still compensating out-of-plane shape anisotropy. Moreover the magnetic signal above 150 K corresponding to Ge$_{3}$Mn$_{5}$ clusters that start to form in this growth temperature range is strongly anisotropic. This perpendicular anisotropy confirms the epitaxial relation: (0002) Ge$_{3}$Mn$_{5}$ $\parallel$ (002) Ge discussed in Ref.\cite{Bihl06}. The magnetic easy axis of the clusters lies along the hexagonal $c$-axis which is perpendicular to the film plane.

\begin{figure}[ht]
\center
   \includegraphics[width=.35\linewidth]{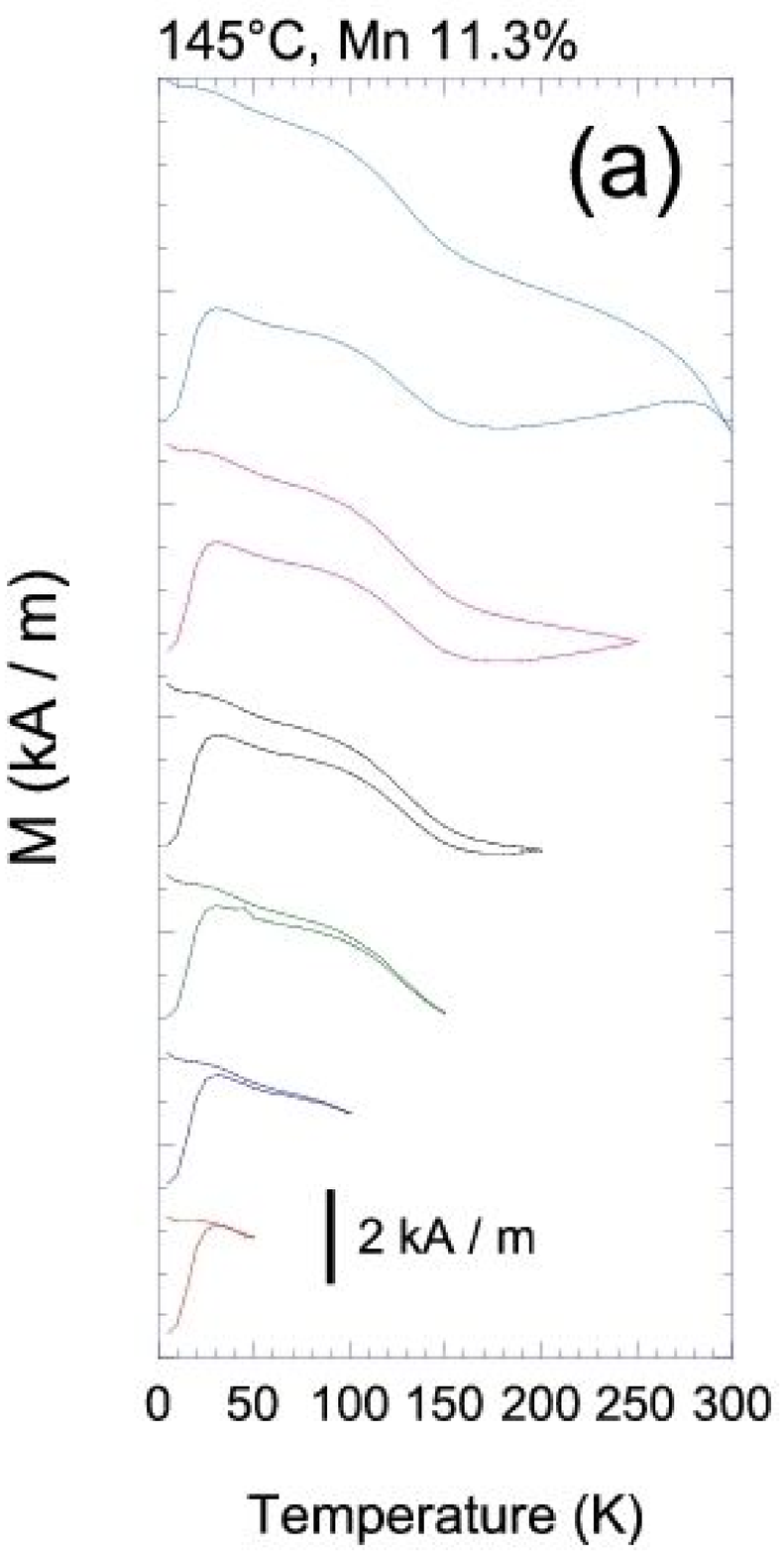}
   \includegraphics[width=.63\linewidth]{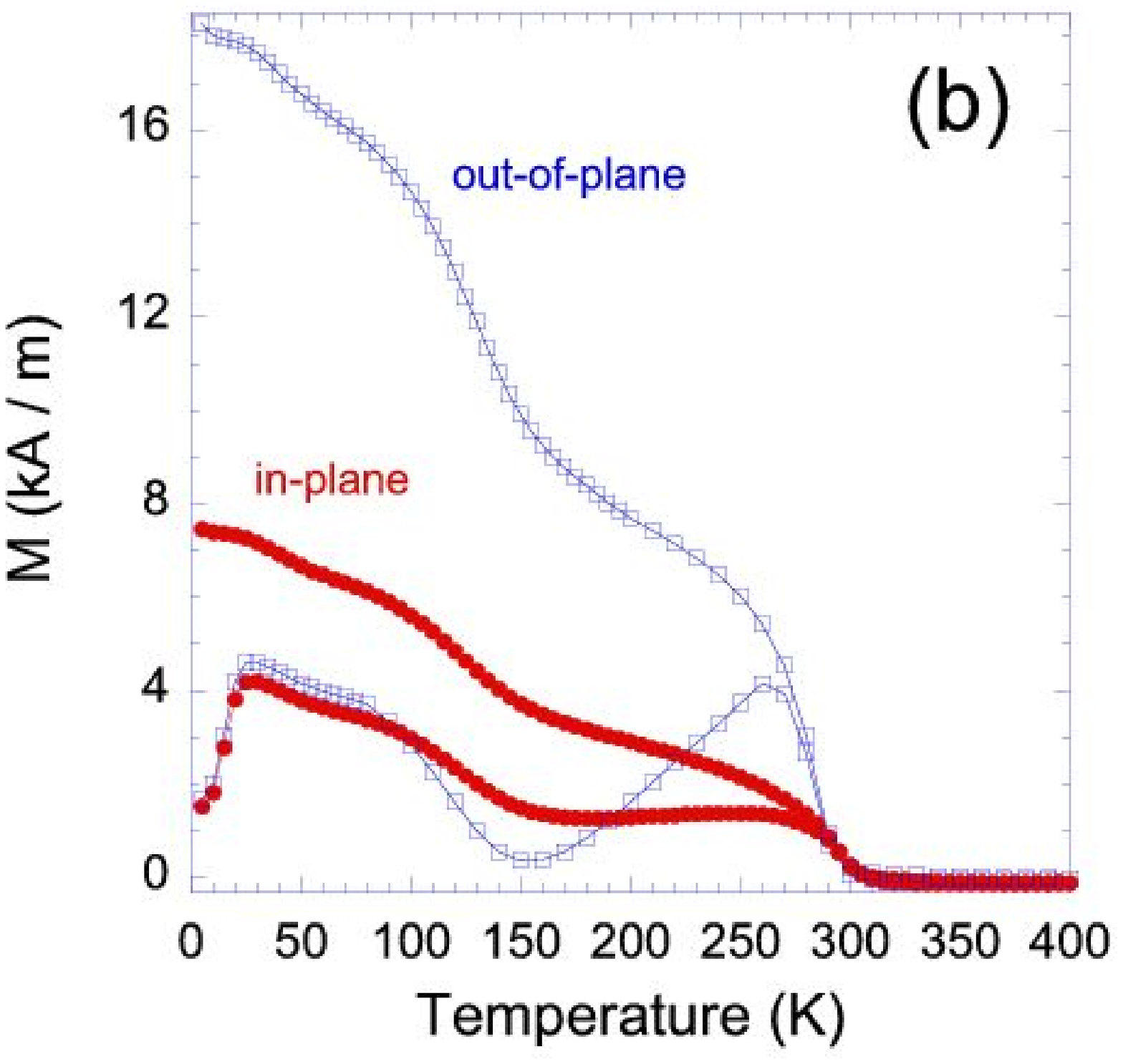} 
\caption{(a) ZFC-FC measurements performed on a Ge$_{0.887}$Mn$_{0.113}$ sample grown at 145$^{\circ}$C. The in-plane applied field is 0.015 T. Magnetization was recorded up to different T$_{0}$ temperatures: 50 K, 100 K, 150 K, 200 K, 250 K and 300 K. Curves are shifted up for more clarity. (b) ZFC-FC curves for in-plane and out-of-plane applied fields (0.015 T).}
\label{fig13}
\end{figure}

For growth temperatures $T_{g}\geq$145$^{\circ}$C the cluster magnetic signal dominates (Fig. 13b). Superparamagnetic nanostructures are investigated performing ZFC-FC measurements at different $T_{0}$ temperatures (Fig. 13a). The first ZFC peak at low temperature \textit{i.e.} $\leq$ 150 K is attributed to  low-$T_{C}$ nanocolumns ($T_{C}\approx$130 K). This peak is wider than for lower growth temperatures and its maximum is further shifted up to 30 K. These results are in agreement with TEM observations: increasing $T_{g}$ leads to larger nanocolumns (\textit{i.e.} higher blocking temperatures) and wider size distributions. ZFC-FC irreversibility is observed up to the Curie temperature due to the presence of ferromagnetic columns. The second peak above 180 K in the ZFC curve is attributed to Ge$_{3}$Mn$_{5}$ clusters and the corresponding ZFC-FC irreversibility persisting up to 300 K means that some clusters are ferromagnetic. We clearly evidence the out-of-plane anisotropy of Ge$_{3}$Mn$_{5}$ clusters and the isotropic magnetic behavior of nanocolumns (Fig. 13b). In this growth temperature range, we have also investigated the Mn concentration dependence of magnetic properties. 

\begin{figure}[ht]
\center
    \includegraphics[width=.49\linewidth]{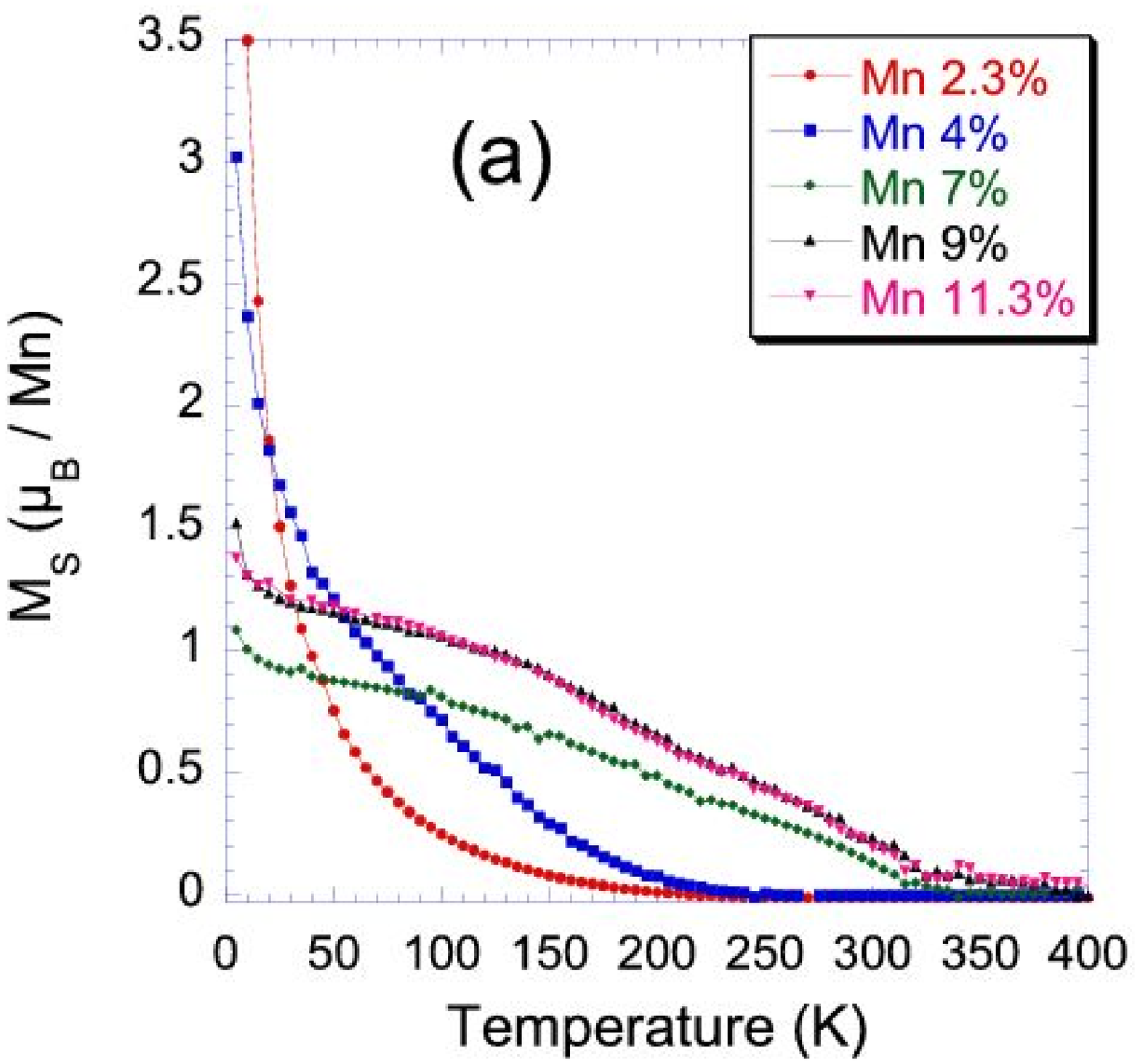}
    \includegraphics[width=.49\linewidth]{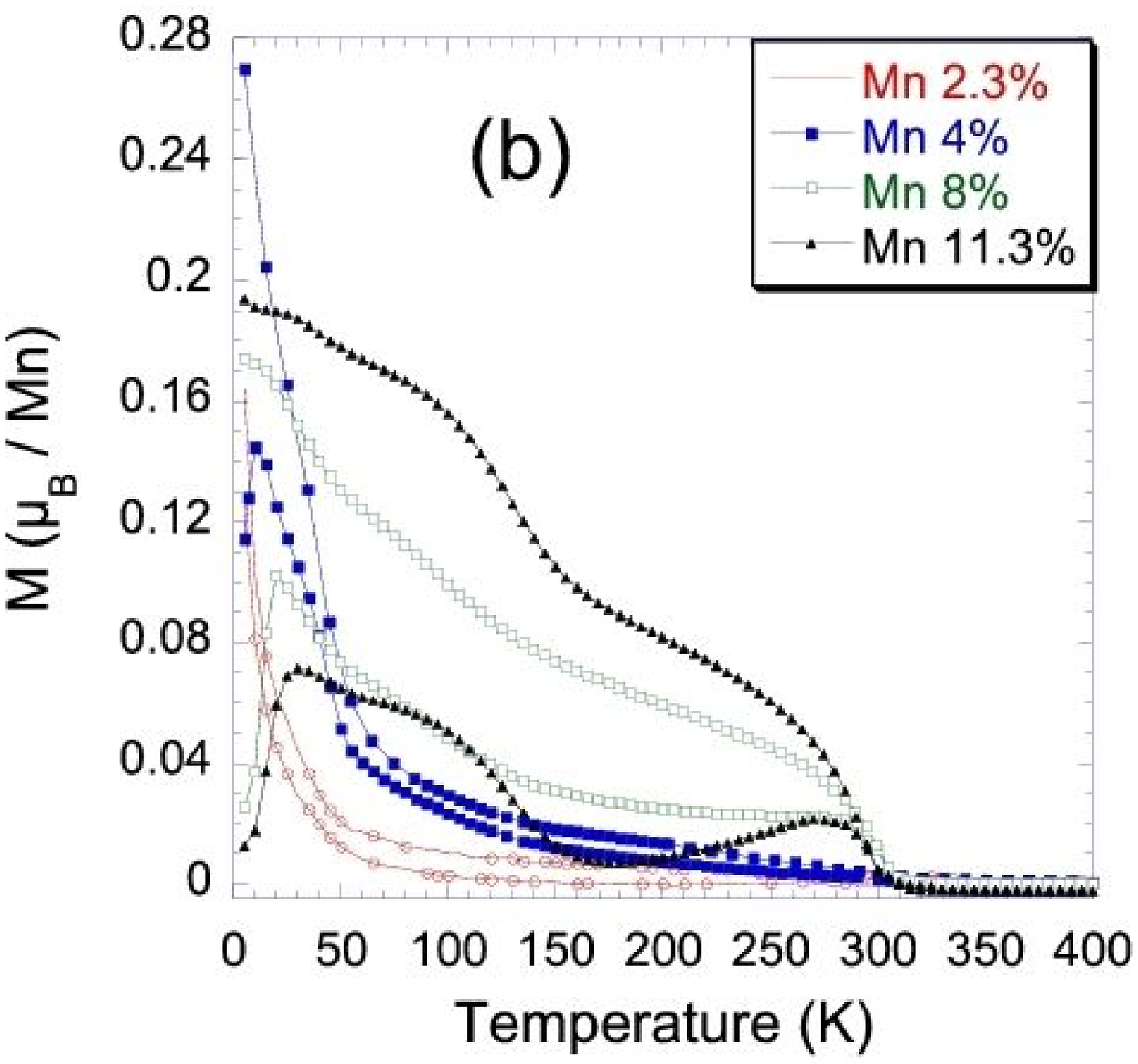} 
\caption{Temperature dependence of the saturation magnetization (in $\mu_{B}$/Mn) of Ge$_{1-x}$Mn$_{x}$ films grown at 150$^{\circ}$C plotted for different Mn concentrations: 2.3 \%; 4 \%; 7 \%; 9 \%; 11.3 \%. (b) ZFC-FC measurements performed on Ge$_{1-x}$Mn$_{x}$ films grown at 150$^{\circ}$C. The in-plane applied field is 0.025 T for 2.3 \% and 4 \% and 0.015 T for 8 \% and 11.3 \%. }
\label{fig14}
\end{figure}

In Fig. 14a, for low Mn concentrations (2.3 \% and 4 \%) the contribution from diluted Mn atoms in the germanium matrix to the saturation magnetization is very high and nearly vanishes for higher Mn concentrations (7 \%, 9 \% and 13 \%) as observed for low growth temperatures. Above 7 \%, the magnetic signal mainly comes from nanocolumns and Ge$_{3}$Mn$_{5}$ clusters. We can derive more information from ZFC-FC measurements (Fig. 14b). Indeed, for 2.3 \% of Mn, ZFC and FC curves nearly superimpose down to low temperature meaning that nanocolumns are superparamagnetic in the whole temperature range. Moreover the weak irreversibility arising at 300 K  means that some Ge$_{3}$Mn$_{5}$ clusters have already formed in the samples even at very low Mn concentrations. For 4 \% of Mn, we can observe a peak with a maximum at the blocking temperature (12 K) in the ZFC curve. We can also derive the Curie temperature of nanocolumns: $\approx$45 K. The irresversibility arising at 300 K still comes from Ge$_{3}$Mn$_{5}$ clusters. Increasing the Mn concentration above 7 \% leads to: higher blocking temperatures (20 K and 30 K) due to larger nanocolumns and wider ZFC peaks due to wider size distributions in agreement with TEM observations (see Fig. 3a). Curie temperatures also increase (110 K and 130 K) as well as the contribution from Ge$_{3}$Mn$_{5}$ clusters.\\
Finally when increasing $T_{g}$ above 160$^{\circ}$C, the nanocolumns magnetic signal vanishes and only Ge$_{3}$Mn$_{5}$ clusters and diluted Mn atoms coexist. The overall magnetic signal becomes comparable to the one measured on annealed samples in which only Ge$_{3}$Mn$_{5}$ clusters are observed by TEM (see Fig. 7a).\\
The magnetic properties of high-$T_{C}$ nanocolumns obtained for $T_{g}$ close to 130$^{\circ}$C are discussed in detail in Ref.\cite{Jame06}.\\
In conclusion, at low growth temperatures ($T_{g}\leq$120$^{\circ}$C), nanocolumns are made of almost 4 independent elongated magnetic nanostructures. For $T_{g}\geq$120$^{\circ}$C, these independent nanostructures percolate into a single nanocolumn sharply leading to higher blocking temperatures. Increasing $T_{g}$ leads to larger columns with a wider size distribution as evidenced by ZFC-FC measurements and given by TEM observations. In parallel, some Ge$_{3}$Mn$_{5}$ clusters start to form and their contribution increases when increasing $T_{g}$. Results on magnetic anisotropy seems counter-intuitive. Indeed Ge$_{3}$Mn$_{5}$ clusters exhibit strong out-of-plane anisotropy whereas nanocolumns which are highly elongated magnetic structures are almost isotropic. This effect is probably due to compensating in-plane magnetoelastic coupling (due to the columns compression) and out-of-plane shape anisotropy. 

\section{Conclusion}

In this paper, we have investigated the structural and magnetic properties of thin Ge$_{1-x}$Mn$_{x}$ films grown by low temperature molecular beam epitaxy. A wide range of growth temperatures and Mn concentrations have been explored. All the samples contain Mn-rich nanocolumns as a consequence of 2D-spinodal decomposition. However their size, crystalline structure and magnetic properties depend on growth temperature and Mn concentration. For low growth temperatures, nanocolumns are very small (their diameter ranges between 1.8 nm for 1.3 \% of Mn and 2.8 nm for 11.3 \% of Mn), their Curie temperature is rather low ($<$ 170 K) and they behave as almost four uncorrelated superparamagnetic nanostructures. Increasing Mn concentration leads to higher columns densities while diameters remain nearly unchanged. For higher growth temperatures, the nanocolumns mean diameter increases and their size distribution widens. Moreover the 4 independent magnetic nanostructures percolate into a single magnetic nanocolumn. Some columns are ferromagnetic even if Curie temperatures remain quite low. In this regime, increasing Mn concentration leads to larger columns while their density remains nearly the same. In parallel, Ge$_{3}$Mn$_{5}$ nanoclusters start to form in the film with their $c$-axis perpendicular to the film plane. In both temperature regimes, the Mn incorporation mechanism in the nanocolumns and/or in the matrix changes above 5 \% of Mn and nanocolumns exhibit an isotropic magnetic behaviour due to the competing effects of out-of-plane shape anisotropy and in-plane magnetoelastic coupling. Finally for a narrow range of growth temperatures around 130$^{\circ}$C, nanocolumns exhibit Curie temperatures higher than 400 K. Our goal is now to investigate the crystalline structure inside the nanocolumns, in particular the position of Mn atoms in the distorted diamond structure, which is essential to understand magnetic and future transport properties in Ge$_{1-x}$Mn$_{x}$ films.

\section{Aknowledgements}
The authors would like to thank Dr. F. Rieutord for grazing incidence x-ray diffraction measurements performed on the GMT station of BM32 beamline at the European Synchrotron Radiation Facility.

\end{document}